\documentclass{iopart}
\usepackage{iopams}  %% for various maths symbols
\expandafter\let\csname equation*\endcsname\relax
\expandafter\let\csname endequation*\endcsname\relax
\usepackage{amsmath} %% incompatible with iopams w/o above two lines
\usepackage{amssymb}
\usepackage{mathtools}
\usepackage{bbold}   %% for a nice identity symbol \mathbb{1} should now work
\usepackage{tensor}  %% for pre-subscripts, see example below
\usepackage{graphicx}  %% for including figure files, example below
\graphicspath{ {FigFiles/} } %% tidy subdir for figure files
\usepackage{subcaption}
\usepackage{natbib}		%% to make references easier
\setcitestyle{square,numbers}	% author year %% use {square,numbers} for numerical citations
\usepackage[breaklinks=true,%
            colorlinks=true,%
            linkcolor=blue,%
            urlcolor=blue,%
            citecolor=blue]{hyperref}
            
\usepackage{circuitikz}
\usepackage{tikz}
\usetikzlibrary{arrows, shapes.gates.logic.US, calc}
\usetikzlibrary{arrows,decorations.pathreplacing}
\usetikzlibrary{backgrounds,fit,decorations.pathreplacing,calc}

\newcommand{\ket}[1]{\ensuremath{\left|#1\right\rangle}}

\usepackage{datetime} % for time stamping current version
\usepackage[textsize=footnotesize,backgroundcolor=red!25]{todonotes}

\usepackage{layouts}

\usepackage[title]{appendix}
\makeatletter
\newrobustcmd{\fixappendix}{%
    \patchcmd{\l@section}{1.5em}{7em}{}{}%
    \patchcmd{\l@subsection}{2.3em}{7em}{}{}%
}
\makeatother

\newcommand{\thedate}{(Dated: $25^{\text{th}}$ January 2021)}

\begin{document}
\title[The controlled SWAP test]{The controlled SWAP test for determining quantum entanglement}
\author{Steph Foulds$^1$, Viv Kendon$^1$, and Tim Spiller$^2$}
\address{$^1$ Department of Physics; Joint Quantum Centre (JQC) Durham-Newcastle, Durham University, South Road, Durham, DH1 3LE, UK}
\address{$^2$ Department of Physics and York Centre for Quantum Technologies, University of York, York YO10 5DD, UK}
\eads{\mailto{stephanie.c.foulds@durham.ac.uk} \\
\thedate}

\begin{abstract}
Quantum entanglement is essential to the development of quantum computation, communications, and technology. The controlled SWAP test, widely used for state comparison, can be adapted to an efficient and useful test for entanglement of a pure state. Here we show that the test can evidence the presence of entanglement (and further, genuine $n$-qubit entanglement), can distinguish entanglement classes, and that the concurrence of a two-qubit state is related to the test's output probabilities. We also propose a multipartite measure of entanglement that acts similarly for $n$-qubit states.
The number of copies required to detect entanglement decreases for larger systems, to four on average for many ($n\gtrsim8$) qubits for maximally entangled states. For non-maximally entangled states, the average number of copies of the test state required to detect entanglement increases with decreasing entanglement.  Furthermore, the results are robust to second order when typical small errors are introduced to the state under investigation.
\end{abstract}

\maketitle

\tableofcontents

%-------------------------------------------------------%
\section{Introduction}\label{sec:intro}
%-------------------------------------------------------%
\markboth{\textit{\rightmark}}{\thepage}

Quantum entanglement is an essential resource for obtaining a quantum advantage in communications \cite{ekert, communication}, metrology \cite{metrology1, metrology}, imaging \cite{imaging1, imaging}, and computation \cite{computation1, computation2, computation}. Quantum teleportation \cite{ogteleport} uses pre-shared entanglement to transfer an unknown quantum state from one location to another using only classical channels. It provides a fundamental primitive for quantum information processing. Teleportation has been realised optically \cite{opticqudits, opticgate} and with ion traps, \cite{iongate} as approaches towards distributed quantum computation \cite{teleportation}.

In general, the amount of entanglement in a state determines its usefulness. For example, information can only be teleported perfectly by maximally entangled states \cite{werner}, which are necessarily pure. The current widely-used method for experimentally determining entanglement, quantum state tomography, does not scale well with an increasing number $n$ of qubits \cite{scalable}. This makes practical alternative tests of interest. In this paper, we investigate a method for detecting entanglement and quantifying the amount of entanglement in a multipartite pure state: the controlled SWAP test.

The controlled SWAP test for entanglement discussed here is an adapted version of the widely used controlled SWAP test for state comparison, which determines whether a pair of states are inequivalent, and requires only a single application of the test to achieve this. This elegance can be applied to the detection and quantification of entanglement in a state, and so is promising as a more efficient alternative to quantum state tomography.
The controlled SWAP test for entanglement was first introduced in \cite{patent}, and then used by Gutowski \textit{et al.}~\cite{markwilde} to prove a series of computational complexity results, by using repeated applications of the controlled SWAP test for entanglement as a product state test.
Although detecting entanglement can be done efficiently, the inverse problem -- determining whether a given state is a separable, i.e., a product state -- is NP-\textsc{hard} \cite{gurvits-separable}.
Our purpose in this paper is practical: to determine the conditions under which the controlled SWAP test for entanglement is likely to be experimentally useful.

The paper is laid out as follows. First, pure state entanglement and its measures are introduced in Section \ref{sec:background}. The controlled SWAP test for state comparison is then explained in Section \ref{sec:equalSWAP}, leading on to its adaptation to the controlled SWAP test for entanglement in Section \ref{sec:entangleSWAP}. Then in Section \ref{sec:ideal}, we present the outcomes of the test for a range of  pure states, and the corresponding results in terms of the multipartite measure of entanglement in Section \ref{sec:degree}. The efficiency of the test for these various states is considered in Section \ref{sec:efficiency}, and finally several typical error scenarios are investigated in Section \ref{sec:errors}.

%----------------------------------------------------------------%
\section{Background}\label{sec:background}

With $\ket{\psi_{1}}$ a normalised superposition of the single qubit computational basis states $\ket{0}$ and $\ket{1}$
\begin{align}
\ket{\psi_1} = A_0 \ket{0} + A_1 \ket{1} \; , \nonumber
\end{align}
with $A_0,A_1\in\mathbb{C}$, the probabilities of measuring outcomes $0$ and $1$ follow respectively from $P(\ket{0}) = {|A_0|}^2$ and $P(\ket{1}) = {|A_1|}^2$. Using the notation $\ket{i}\ket{j} = \ket{ij}$ with $i,j\in \{0,1\}$ for the basis states of multiple qubit systems, a general two-qubit state takes the form \cite{nielsen}
\begin{align}\label{super2}
\ket{\psi_{2}} = A_{00} \ket{00} + A_{01} \ket{01} + A_{10} \ket{10} + A_{11} \ket{11} , 
\end{align}
with $A_{ij}\in\mathbb{C}$ and normalisation $\sum_{i,j}|A_{ij}|^2=1$.

A composite system $\ket{\psi_{2}}$ is in an entangled state if it cannot be written as a product state for its component systems, i.e. $\ket{\psi_{2}} \neq \ket{\psi_1} \ket{\phi_1}$ for any pure states $\ket{\psi_1}$, $\ket{\phi_1}$.
The concurrence  \cite{wootters}
\begin{align} \label{c}
C_2 = 2 |A_{00} A_{11} - A_{01} A_{10}|
\end{align}
is a measure of two-qubit entanglement, with $0 \leq C_2 \leq 1$, so a separable or `product' state has $C_2 = 0$ and a maximally entangled state has $C_2 = 1$.  There are many other measures of entanglement which quantify the amount of entanglement in a state, though for bipartite pure states they can all be shown to be equivalent to the entropy of the subsystems \cite{Popescu}.

For two qubit states, concurrence is simpler to calculate, so more convenient for our work. More generally, an entanglement measure has to satisfy certain properties, such as not increasing on average under local operations and classical communication (LOCC) \cite{nielsen}. In essence, it must be true that \cite{Emeasure}
\begin{align} \label{entanglementdegree}
    \varepsilon (\rho) \geq \sum_{j=1} p_j \varepsilon(\rho_j) ,
\end{align}
where $\varepsilon$ is an entanglement measure, $j$ refers to the outcomes of a local measurement, and $p_j$ the probability of these outcomes.

For two qubit entangled states, the possibilities are straightforward.
There are four orthogonal maximally entangled two-qubit states \cite{nielsen,nature}, known as the Bell states:
\begin{align} \label{Bell}
\ket{\Phi^{\pm}} &= \frac{\ket{00} \pm \ket{11}}{\sqrt{2}} , \nonumber\\
\ket{\Psi^{\pm}} &= \frac{\ket{01} \pm \ket{10}}{\sqrt{2}} .
\end{align}
Under reversible LOCC, Bell states can be transformed into one another, but cannot be transformed into a state with less than maximal entanglement. Bell states are therefore considered equivalent to one another and form a  unique class of maximally entangled two-qubit states.

For systems with more than two qubits, there are multiple distinct classes of entanglement, based on whether the states can be transformed into each other under reversible LOCC.  Within each class, there is a subset of maximally entangled states.
For example, for three qubits, there are two types of pure entangled states, GHZ \cite{GHZW} and W states, which are inequivalent under reversible LOCC.
For more qubits, GHZ and W states naturally generalise, alongside further distinct types of entangled states.  
In the computational basis for $n$ qubits where $n\geq3$, the maximally entangled GHZ and W states are
\cite{class}
\begin{align}
    \ket{\text{GHZ}_n} &= \frac{1}{\sqrt{2}} (\ket{0}^n + \ket{1}^n) , \label{defGHZ} \\
    \ket{\text{W}_n} &= \frac{1}{\sqrt{n}} \sum^n_{i=1} \ket{0...1_i...0} \; , \label{defW}
\end{align}
where we have introduced the notation $\ket{0}^n$ for $n$ qubits all in state $\ket{0}$.
Under reversible LOCC, GHZ and W states remain maximally entangled within their respective classes.
However, GHZ states are considered more entangled than W states, based on various pure state entanglement measures; while W states are more robust, as loss or measurement of some qubits can still leave an entangled state of the remainder \cite{class}.
In this paper, we focus on these two types of entangled states, which are of particular interest due to their applications in quantum computing \cite{GHZW,GHZapp, Wapp}.  

Reversible (unitary) transformations of quantum states can be represented as quantum gates, the model used in this paper. An important and relevant single-qubit example is the Hadamard gate H: \cite{nielsen}
\begin{align} \label{hadamard}
     \text{H} \ket{0} = \frac{1}{\sqrt{2}} \left( \ket{0} + \ket{1} \right) \quad \text{and} \quad
     \text{H} \ket{1} = \frac{1}{\sqrt{2}} \left( \ket{0} - \ket{1} \right) .
\end{align}
Multi-qubit gates of relevance include the two-qubit CNOT gate which flips the target qubit only if the control qubit is $\ket{1}$. It can be represented by the matrix
\begin{align} \label{CNOT}
\text{CNOT} = 
\begin{bmatrix}
1 & 0 & 0 & 0\\
0 & 1 & 0 & 0\\
0 & 0 & 0 & 1\\
0 & 0 & 1 & 0
\end{bmatrix}
\end{align}
where the first qubit is the control and the second the target.
The three-qubit Toffoli gate has two controls and one target: the target qubit is flipped only if both the controls are $\ket{1}$. It has the matrix
\begin{align} \label{Toffoli}
\text{T} = 
\begin{bmatrix}
1 & 0 & 0 & 0 & 0 & 0 & 0 & 0\\
0 & 1 & 0 & 0 & 0 & 0 & 0 & 0\\
0 & 0 & 1 & 0 & 0 & 0 & 0 & 0\\
0 & 0 & 0 & 1 & 0 & 0 & 0 & 0\\
0 & 0 & 0 & 0 & 1 & 0 & 0 & 0\\
0 & 0 & 0 & 0 & 0 & 1 & 0 & 0\\
0 & 0 & 0 & 0 & 0 & 0 & 0 & 1\\
0 & 0 & 0 & 0 & 0 & 0 & 1 & 0\\
\end{bmatrix}
\end{align}
where the first two qubits are the controls.

%%%%%%%%%%%%%%%%%%%%%%%%%%%%%%%%%%%%%%%%%%%%%%
\subsection{Experimentally determining entanglement}\label{sec:previous}

The most commonly used method for determining entanglement is quantum state tomography. Quantum state tomography builds a system's density matrix entry by entry in order to derive its entanglement. A large ensemble of identical states is prepared to carry out the required number of measurements \cite{tomog,matrix,tomog2}. The density matrix grows exponentially with the system size and so this method becomes unfavourable for large systems. For an $n$-qubit state, the number of measurements required is typically \cite{scalable,tomog2} in the order of $3^n$.

If only the entanglement of the system is of interest, there are alternative methods that are arguably more efficient. Entanglement witnesses are functionals of a state's density matrix that can be directly measured and determine whether a state is entangled. To obtain the witness of an $n$-qubit state, as few as $2n - 1$ measurements are required. However, the witness must be optimised for the state, and so this is not a general method \cite{twon,nature2,construction}. 

Many attempts have been made to improve upon the above methods in terms of efficiency and generality.
The experiment in Walborn \textit{et al.}~\cite{nature} determines how entangled two-qubit states are by measuring only the final polarisation. This method is able to detect and distinguish Bell states; the probability of measuring the $\ket{\Psi^-}$ Bell state is then related to the concurrence to quantify the entanglement. Thousands of measurements are needed to achieve sufficient statistics. Theory extending this method to any number of qubits is provided by Harrow and Montanaro \cite{Merlin}. Further proposals include Ekert \textit{et al.}~\cite{eigen} which is also based on the controlled SWAP test for state comparison, and Amaro \textit{et al.}~\cite{betterwitness} which is an improved witness-based method.

%%%%%%%%%%%%%%%%%%%%%%%%%%%%%%%%%%%%%%%%%%%%%%
\subsection{The controlled SWAP test for equivalence}\label{sec:equalSWAP}

\begin{figure}[t]
\begin{subfigure}[b]{0.5\textwidth}
    \centering
    \begin{tikzpicture}[thick]
        \tikzset{operator/.style = {draw,fill=white,minimum size=1.5em}, operator2/.style = {draw,fill=white,minimum height=2cm, minimum width=1cm}, phase/.style = {draw,fill,shape=circle,minimum size=5pt,inner sep=0pt}, surround/.style = {fill=blue!10,thick,draw=black,rounded corners=2mm}, cross/.style={path picture={\draw[thick,black](path picture bounding box.north) -- (path picture bounding box.south) (path picture bounding box.west) -- (path picture bounding box.east); }}, crossx/.style={path picture={ \draw[thick,black,inner sep=0pt] (path picture bounding box.south east) -- (path picture bounding box.north west) (path picture bounding box.south west) -- (path picture bounding box.north east); }}, circlewc/.style={draw,circle,cross,minimum width=0.3 cm}, meter/.style= {draw, fill=white, inner sep=7, rectangle, font=\vphantom{A}, minimum width=25, line width=.8, path picture={\draw[black] ([shift={(.1,.3)}]path picture bounding box.south west) to[bend left=50] ([shift={(-.1,.3)}]path picture bounding box.south east);\draw[black,-latex] ([shift={(0,.1)}]path picture bounding box.south) -- ([shift={(.3,-.1)}]path picture bounding box.north);}}, } \matrix[row sep=0.4cm, column sep=0.8cm] (circuit) {\node (q1) {$\ket{\psi}_A$}; & & \node[](P21){}; & &\coordinate (end1);\\ \node (q2) {$\ket{\phi}_B$}; & & & &\coordinate (end2);\\ \node (q3) {$\ket{0}_C$}; & \node[operator] (H11) {H}; & \node[phase] (P11) {}; & \node[operator] (H12) {H}; & & [-0.8cm] \node[meter] (meter) {}; \coordinate (end3); \\ }; \begin{pgfonlayer}{background} \draw[thick] (q1) -- (end1) (q2) -- (end2) (q3) -- (end3) (P11) -- (P21); \draw[thick,fill=white] (1.2,-0.1) rectangle (-1.4,1.3) node [pos=.5]{SWAP}; \end{pgfonlayer}
    \end{tikzpicture}
    \caption{\label{1qubitcswap} }
\end{subfigure}\hfill
\begin{subfigure}[b]{0.5\textwidth}
    \centering
    \begin{tikzpicture}[thick]
        \tikzset{operator/.style = {draw,fill=white,minimum size=1.5em}, operator2/.style = {draw,fill=white,minimum height=2cm, minimum width=1cm}, phase/.style = {draw,fill,shape=circle,minimum size=5pt,inner sep=0pt}, phase2/.style = {draw,fill=blue,shape=circle,minimum size=5pt,inner sep=0pt}, phase3/.style = {draw,fill=red,shape=circle,minimum size=5pt,inner sep=0pt}, surround/.style = {fill=blue!10,thick,draw=black,rounded corners=2mm}, cross/.style={path picture={ \draw[thick,black](path picture bounding box.north) -- (path picture bounding box.south) (path picture bounding box.west) -- (path picture bounding box.east);}}, circlewc/.style={draw,circle,cross,minimum width=0.3 cm}, cross2/.style={path picture={ \draw[thick,blue](path picture bounding box.north) -- (path picture bounding box.south) (path picture bounding box.west) -- (path picture bounding box.east);}}, circlewc2/.style={draw,color=blue,circle,cross2,minimum width=0.3 cm}, cross3/.style={path picture={ \draw[thick,red](path picture bounding box.north) -- (path picture bounding box.south) (path picture bounding box.west) -- (path picture bounding box.east); }}, circlewc3/.style={draw,color=red,circle,cross3,minimum width=0.3 cm}, meter/.style= {draw, fill=white, inner sep=7, rectangle, font=\vphantom{A}, minimum width=25, line width=.8, path picture={\draw[black] ([shift={(.1,.3)}]path picture bounding box.south west) to[bend left=50] ([shift={(-.1,.3)}]path picture bounding box.south east);\draw[black,-latex] ([shift={(0,.1)}]path picture bounding box.south) -- ([shift={(.3,-.1)}]path picture bounding box.north);}},} \matrix[row sep=0.4cm, column sep=0.8cm] (circuit) {\node (q1) {$\ket{\psi}_A$}; &\node[phase2] (P11) {}; &\node[circlewc3] (P12) {}; &\node[phase2] (P13) {}; &\coordinate (end1);\\ \node (q2) {$\ket{\phi}_B$}; &\node[circlewc2] (P21) {}; &\node[phase3] (P22) {}; &\node[circlewc2] (P23) {}; &\coordinate (end2);\\ \node (q3) {$\ket{0}_C$}; &\node[operator] (H11) {H}; &\node[phase3] (P32) {}; &\node[operator] (H12) {H}; & &[-0.8cm] \node[meter] (meter) {}; \coordinate (end3); \\};
        \begin{pgfonlayer}{background} \draw[thick] (q1) -- (end1) (q2) -- (end2) (q3) -- (end3); \draw[thick, blue] (P11) -- (P21) (P13) -- (P23); \draw[thick, red] (P12) -- (P22) -- (P32); \end{pgfonlayer}
    \end{tikzpicture}
    \caption{\label{swapgate}}
\end{subfigure}\hfill
\caption{The quantum circuit for an equivalency SWAP test on the two states $\ket{\psi}$ and $\ket{\phi}$. H is a Hadamard gate from equation \eqref{hadamard}. (a) The SWAP gate swaps all qubits in the test states on the condition that the control qubit is in state $\ket{1}$. (b) shows the SWAP gate broken down into individual gates for the one-qubit test state case. The central gate, shown in red, is a Toffoli gate from equation \eqref{Toffoli} and the two gates either side in blue are CNOT gates from equation \eqref{CNOT}, where the crossed circles are controlled on the dots. The final CNOT gate -- not necessary for the test outcome -- returns the system to its initial state in the case of equivalent states. \label{equiv}}
\end{figure}
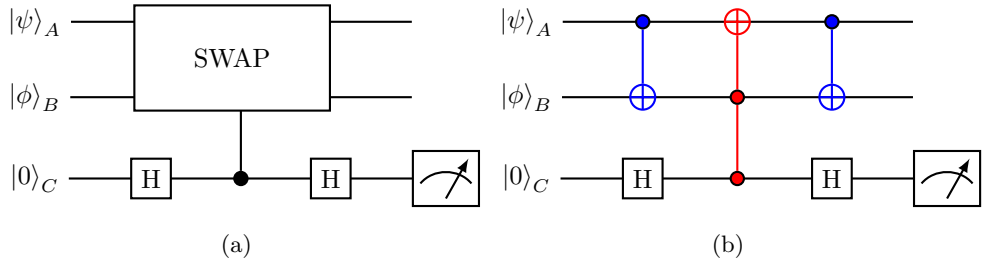

The controlled SWAP test for equivalence is a widely applied method for determining whether two given pure $n$-qubit states $\ket{\psi}$ and $\ket{\phi}$ are equivalent, detailed in \cite{originalCSWAP} and its optical implementation in \cite{patent}. The circuit for this procedure can be seen in Figure \ref{1qubitcswap}. Three states are required, with the initial composite state $\ket{\Psi} = \ket{\psi}_A \ket{\phi}_B \ket{0}_C$. A Hadamard gate is applied to the control qubit $C$, followed by a controlled-SWAP gate on the two test states $A$ and $B$, controlled on the single qubit \cite{patent, number, digital}.
The SWAP gate is applied according to the state at the control qubit $C$: if $\ket{C} = \ket{0}_C$ there is no change, whereas $\ket{1}_C$ will result in the states of $A$ and $B$ being swapped \cite{number}. In the case of a single qubit state comparison, the SWAP gate is composed of two CNOT gates \cite{nielsen} and a Toffoli gate \cite{nielsen}, as shown in Figure \ref{swapgate} \cite{patent,implementation}.
Finally, another Hadamard gate is applied to the control qubit. The resulting composite state is then
\begin{align}
\ket{\Psi} = \frac{1}{2} [ (\ket{\phi}_A \ket{\psi}_B + \ket{\psi}_A \ket{\phi}_B) \ket{0}_C + (\ket{\phi}_A \ket{\psi}_B - \ket{\psi}_A \ket{\phi}_B) \ket{1}_C ] .
\nonumber \end{align}
It is clear that if $\ket{\phi} = \ket{\psi}$, the control qubit will be in $\ket{0}_C$ with absolute certainty. Measuring the control qubit in $\ket{1}_C$ therefore proves that the two states $A$ and $B$ are inequivalent. If the two states are identical, confidence of this outcome is increased by repeating the test and obtaining multiple measurements of $\ket{0}_C$ with no measurements of $\ket{1}_C$ \cite{number, patent, digital}.

%----------------------------------------------------------------%
\section{The controlled SWAP test for entanglement}\label{sec:entangleSWAP}

The controlled SWAP (c-SWAP) test for state comparison can be modified to instead test for entanglement. This is outlined for the two-qubit state case in van Dam \textit{et al.}~\cite{patent}, along with a potential optical setup.
Harrow and Montanaro \cite{merlinarthur} discuss multiple applications of the c-SWAP test as a product-state test, proving its correctness for all numbers of qubits and its optimal soundness. Following this, Gutoski \textit{et al.}~\cite{markwilde} prove the product-state c-SWAP test is a complete problem for the complexity class BQP. This section details the theory of the c-SWAP test for entanglement.

\begin{figure}[t]
\begin{subfigure}[b]{0.5\textwidth}
    \centering
    \begin{tikzpicture}[thick]
        \tikzset{
        operator/.style = {draw,fill=white,minimum size=0.1em}, 
        operator2/.style = {draw,fill=white,minimum height=2cm, minimum width=1cm}, 
        phase/.style = {draw,fill,shape=circle,minimum size=5pt,inner sep=0pt}, 
        surround/.style = {fill=blue!10,thick,draw=black,rounded corners=2mm}, 
        cross/.style={path picture={\draw[thick,black](path picture bounding box.north) -- (path picture bounding box.south) (path picture bounding box.west) -- (path picture bounding box.east); }}, crossx/.style={path picture={ \draw[thick,black,inner sep=0pt] (path picture bounding box.south east) -- (path picture bounding box.north west) (path picture bounding box.south west) -- (path picture bounding box.north east); }}, 
        circlewc/.style={draw,circle,cross,minimum width=0.3 cm}, 
        meter/.style= {draw, fill=white, inner sep=5, rectangle, font=\vphantom{A}, minimum width=20, line width=.8, path picture={\draw[black] ([shift={(.1,.2)}]path picture bounding box.south west) to[bend left=30] ([shift={(-.1,.2)}]path picture bounding box.south east);\draw[black,-latex] ([shift={(0,.1)}]path picture bounding box.south) -- ([shift={(.2,-.1)}]path picture bounding box.north);}}, }
        \matrix[row sep=0.2cm, column sep=0.35cm] (circuit) {
        \node (q1) {}; % LINE a1
        &\node[phase] (P10) {}; & & \node[circlewc] (P1) {}; & & \node[phase] (P11) {}; & \coordinate (end1);
        \\ \node (q2) {}; % LINE a2
        && \node[phase] (P14) {}; & & \node[circlewc] (P2) {}; & & \node[phase] (P12) {}; \coordinate (end2);
        \\ \node (q3) {}; % LINE b1
        &\node[circlewc] (P3) {}; & & \node[phase] (P4) {}; & & \node[circlewc] (P5) {}; & \coordinate (end3);
        \\ \node (q4) {}; % LINE b2
        && \node[circlewc] (P13) {}; & & \node[phase] (P6) {}; & & \node[circlewc] (P7) {};  \coordinate (end4);
        \\ \node (q5) {$\ket{0}_C$}; % LINE c1
        && \node[operator] (H11) {H};
        & \node[phase] (P8) {}; & ;
        & \node[operator] (H12) {H};
        & & [-0.8cm] \node[meter] (meter) {}; \coordinate (end5);
        \\ \node (q6) {$\ket{0}_C$}; % LIVE c2
        && \node[operator] (H11) {H}; & ;
        & \node[phase] (P9) {};
        & \node[operator] (H12) {H};
        & & [-0.8cm] \node[meter] (meter) {}; \coordinate (end6); \\ };
        \draw [decorate,decoration={brace,amplitude=5pt},xshift=-0.5cm,yshift=0pt]
(q2) -- (q1) node [black,midway,xshift=-0.6cm]
{\footnotesize $\ket{A}$};
        \draw [decorate,decoration={brace,amplitude=5pt},xshift=-0.5cm,yshift=0pt]
(q4) -- (q3) node [black,midway,xshift=-0.6cm]
{\footnotesize $\ket{B}$};
        \begin{pgfonlayer}{background} 
        \draw[thick] (q1) -- (end1) (q2) -- (end2) (q3) -- (end3) (q4) -- (end4) (q5) -- (end5) (q6) -- (end6) (P1) -- (P8) (P2) -- (P9) (P10) -- (P3) (P5) -- (P11) (P7) -- (P12) (P13) -- (P14) ; \end{pgfonlayer}
    \end{tikzpicture}
    \caption{\label{2a}}
\end{subfigure}
\begin{subfigure}[b]{0.5\textwidth}
    \centering
    \begin{tikzpicture}[thick]
        \tikzset{operator/.style = {draw,fill=white,minimum size=1.5em}, operator2/.style = {draw,fill=white,minimum height=2cm, minimum width=1cm}, phase/.style = {draw,fill,shape=circle,minimum size=5pt,inner sep=0pt}, surround/.style = {fill=blue!10,thick,draw=black,rounded corners=2mm}, cross/.style={path picture={\draw[thick,black](path picture bounding box.north) -- (path picture bounding box.south) (path picture bounding box.west) -- (path picture bounding box.east); }}, crossx/.style={path picture={ \draw[thick,black,inner sep=0pt] (path picture bounding box.south east) -- (path picture bounding box.north west) (path picture bounding box.south west) -- (path picture bounding box.north east); }}, circlewc/.style={draw,circle,cross,minimum width=0.3 cm}, meter/.style= {draw, fill=white, inner sep=7, rectangle, font=\vphantom{A}, minimum width=25, line width=.8, path picture={\draw[black] ([shift={(.1,.3)}]path picture bounding box.south west) to[bend left=50] ([shift={(-.1,.3)}]path picture bounding box.south east);\draw[black,-latex] ([shift={(0,.1)}]path picture bounding box.south) -- ([shift={(.3,-.1)}]path picture bounding box.north);}}, } \matrix[row sep=0.4cm, column sep=0.8cm] (circuit) {\node (q1) {\ket{A}}; & & \node[](P21){}; & &\coordinate (end1);\\ \node (q2) {\ket{B}}; & & & &\coordinate (end2);\\ \node (q3) {\ket{C}}; & \node[operator] (H11) {H}; & \node[phase] (P11) {}; & \node[operator] (H12) {H}; & & [-0.8cm] \node[meter] (meter) {}; \coordinate (end3); \\ }; \begin{pgfonlayer}{background} \draw[thick] (q1) -- (end1) (q2) -- (end2) (q3) -- (end3) (P11) -- (P21); \draw[thick,fill=white] (1.2,-0.1) rectangle (-1.4,1.3) node [pos=.5]{SWAP}; \node[] at (1.49,1.29) {$\otimes n$}; \node[] at (-0.72,-0.75) {$\otimes n$}; \node[] at (1.66,-0.75) {$\otimes n$}; \end{pgfonlayer}
    \end{tikzpicture}
    \caption{\label{2b}}
\end{subfigure}
\caption{The quantum circuit used to carry out a SWAP test for entanglement on test state $\ket{A}$ and copy state $\ket{B}$. H denotes a Hadamard gate. Initially, $\ket{C} = \ket{0}_C^n$. (a) shows the SWAP gate broken down into individual gates in the case of a two-qubit test state, composed of CNOT gates and Toffoli gates. The final two CNOT gates are to return the test and copy states to their original states (in some cases) and so are optional. (b) shows the circuit for an $n$-qubit test state in compact form. \label{qubitcswap}}
\end{figure}
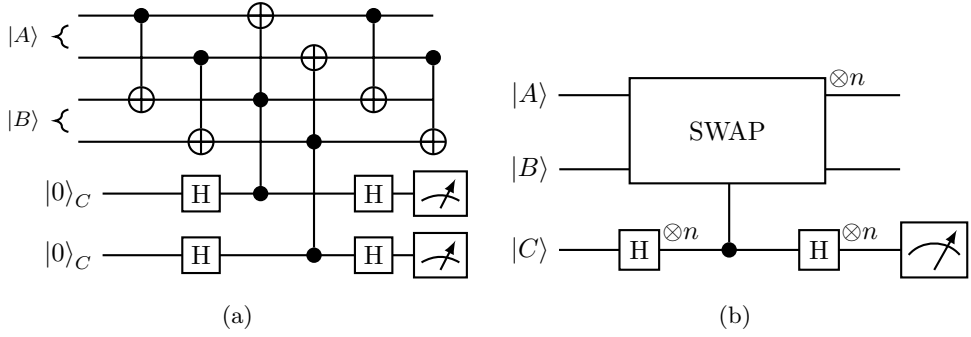

The quantum circuit used for state comparison, Figure \ref{equiv}, is adapted to Figure \ref{qubitcswap}. Two copies of the state to be tested for entanglement are required, labelled $\ket{A}_A$ and $\ket{B}_B$, and several control qubits -- one control qubit for each  qubit in the test state. Figure \ref{2a} shows the quantum circuit for the simplest case, a two qubit entangled state.  Initially, the control state is in $\ket{00}_C$. The two Hadamard gates and the SWAP gate act on each qubit in the control state. The SWAP gate is applied to the corresponding qubits in the test states such that the $i$th qubits in the test and copy states are swapped with one another if the $i$th qubit in the control is $\ket{1}_C$.
The initial composite state is
\begin{align}
\ket{\Psi} = \ket{A}_A \ket{B}_B \ket{00}_C .
\nonumber \end{align}
Passing this system through the entire test in Figure \ref{qubitcswap} gives the final result
\begin{align}
\ket{\Psi} = \frac{1}{4} \sum_{ijrs} \ket{ij}_A \ket{rs}_B 
[ (A_{ij} B_{rs} + A_{is} B_{rj} + A_{rj} B_{is} + A_{rs} B_{ij}) &\ket{00}_C, \nonumber\\
+ (A_{ij} B_{rs} - A_{is} B_{rj} + A_{rj} B_{is} - A_{rs} B_{ij}) &\ket{01}_C, \nonumber\\
+ (A_{ij} B_{rs} + A_{is} B_{rj} - A_{rj} B_{is} - A_{rs} B_{ij}) &\ket{10}_C, \nonumber\\
+ (A_{ij} B_{rs} - A_{is} B_{rj} - A_{rj} B_{is} + A_{rs} B_{ij}) &\ket{11}_C ].
\nonumber \end{align}
Ideally, the copy state is an exact copy of the test state. In this case, the above equation reduces to
\begin{align}
\ket{\Psi} = \frac{1}{2} \sum_{ijrs} \ket{ij}_A \ket{rs}_B [(A_{ij} A_{rs} + A_{is} A_{rj}) &\ket{00}_C \nonumber\\
+ (A_{ij} A_{rs} - A_{is} A_{rj}) &\ket{11}_C ]
\nonumber \end{align}
and so the probability of the control being in $\ket{01}_C$ or $\ket{01}_C$ is zero. This expression can in fact be written in terms of the concurrence $C_2$ (as given by equation \eqref{c}):
\begin{align} \label{finalstate}
    \ket{\Psi} = \text{\Large{[}}& \ket{A}_A \ket{A}_B \nonumber\\
    &\pm \frac{1}{2} C_2 \cdot \frac{1}{2} (-\ket{00}_A \ket{11}_B + \ket{01}_A \ket{10}_B + \ket{10}_A \ket{01}_B - \ket{11}_A \ket{00}_B) \text{\Large{]}} \ket{00}_C \nonumber\\
    &\pm \frac{1}{2} C_2 \cdot \frac{1}{2} (\ket{00}_A \ket{11}_B - \ket{01}_A \ket{10}_B - \ket{10}_A \ket{01}_B + \ket{11}_A \ket{00}_B) \ket{11}_C
\end{align}
with the $\pm$s being $+$ in the case that $A_{00}A_{11}>A_{01}A_{10}$ and $-$ if $A_{00}A_{11}<A_{01}A_{10}$.
If the system is in a product state then $C_2=0$; applying this to the above equation gives
\begin{align}
\ket{\Psi} = \ket{A}_A \ket{A}_B \ket{00}_C
\end{align}
and so the control state is $\ket{00}_C$ with certainty. Any measurement of $\ket{11}$ for the control qubits therefore proves a non-zero concurrence, and evidences the presence of entanglement \cite{patent} in state $\ket{A}$.

Note that if the test state is a product state, and only then, the final state is the same as the initial state. In this case, the test is non-destructive and so the output state can be used as an input state in the next test iteration.

To expand the setup to any number of qubits $n$, the test simply requires $n$ control qubits, as shown in Figure \ref{2b}. The summation notation above as derived by \cite{patent} demonstrates that certain outcomes of the control state evidence entanglement in the test state. However, fully investigating the capability of the c-SWAP test requires the derivation of the expanded resulting state, which we discuss in the next section.

%------------------------------------------------------------------------%
\section{The controlled SWAP test on ideal states}\label{sec:results}

We have analytically derived the final probability distributions in the control state for the most general pure two-qubit and three-qubit test states.  The final expressions are quite lengthy, and are given in \ref{2full} and \ref{3full} respectively. 
Specialising to more symmetric classes of states, for which these expressions simplify, we then extrapolated the probability expressions to those for $n$-qubit states, verified these by computation up to six qubits for non-symmetric test states, and to eight qubits for symmetric test states. In this section, we present the results for GHZ and W states, discuss their relationship to the amount of entanglement in the test state, and investigate the measurement efficiency of the test.

%%%%%%%%%%%%%%%%%%%%%%%%%%%%%%%%%%%%%%%%%%%%%%
\subsection{Bell, GHZ, and W states}\label{sec:ideal}

If the test state is a product state, then only $\ket{0}$ will be measured for any qubit in the control state. In the ideal case, where the copy state is an exact copy of the test state (unequal copy states are investigated in section \ref{sec:unequal}), a measurement of any number of $\ket{1}$s in the control evidences the presence of entanglement. These states, with one or more $\ket{1}$s, that provide evidence of entanglement we call \emph{entanglement signatures}.

% 2-qubit
If the test state is a Bell state and the copy state is an exact copy, for example $\ket{A}=\ket{B}=\ket{\Psi^+}$ from equation \eqref{Bell}, the resulting probability distribution in the control state is:
\begin{align}
    P(\ket{00}_C) &= \frac{3}{4}, \nonumber\\
    P(\ket{01}_C) &= 0, \nonumber\\
    P(\ket{10}_C) &= 0, \nonumber\\
    P(\ket{11}_C) &= \frac{1}{4} .
\end{align}
As seen in Section \ref{sec:entangleSWAP}, any measurement of $\ket{11}_C$ evidences some amount of entanglement for any two-qubit state. Therefore this is the entanglement signature for all two-qubit systems. 

% n-qubit
When $\ket{A} = \ket{B} = \ket{\text{GHZ}_n}$ as in equation \eqref{defGHZ}, the probability results in the control state are
\begin{align}
    P(\ket{0}^n_C) &= \frac{1}{2} + \frac{1}{2^n}, \nonumber\\
    P(\ket{\text{even no. of 1s}}_C)
    &= \frac{1}{2}-\frac{1}{2^n}
\end{align}
where $n$ is the number of qubits in the test state. All other states have zero probability. The states $\ket{\text{even no. of 1s}}_C$ are the entanglement signatures for any GHZ-like state.

When $\ket{A} = \ket{B} = \ket{\text{W}_n}$ as in equation \eqref{defW}, the probability expressions in terms of $n$ are
\begin{align}
    P(\ket{0}^n_C) &= \frac{1}{2} + \frac{1}{2n}, \nonumber\\
    P(\ket{\text{exactly two 1s}}_C) &= \frac{1}{2} - \frac{1}{2n}
\end{align}
with $\ket{\text{exactly two 1s}}_C$ as the entanglement signatures for W-like states.
For $n=2$, the equations for both GHZ and W reduce to those for Bell states, and so for much of the paper we only consider Bell states as the GHZ/W $n=2$ case.

\begin{figure}[t]
    \centering
    \includegraphics[width=.7\linewidth]{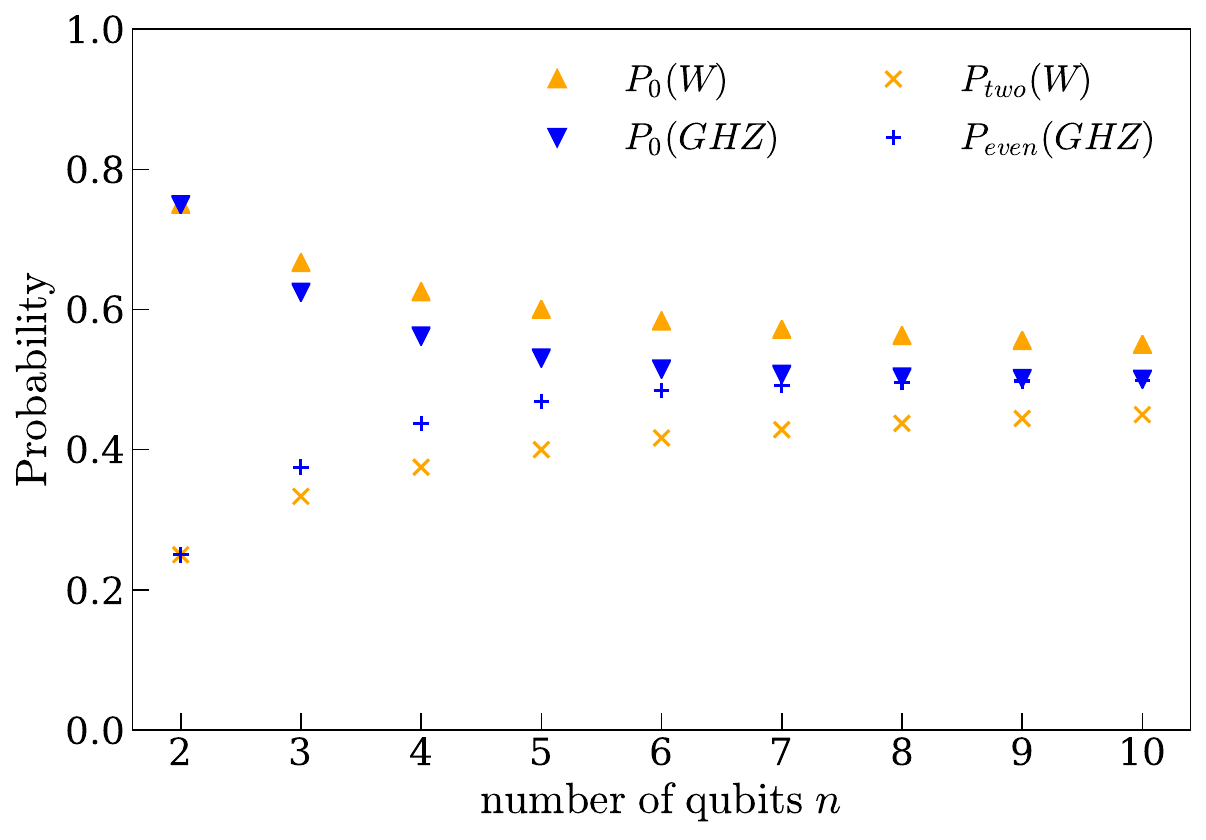}
    \caption{The probability results for maximally entangled GHZ and W states for increasing number of qubits $n$. $P_0$ refers to $P(\ket{0}^n_C)$, $P_{even}$ to $P(\ket{\text{even no. of 1s}}^n_C)$, and $P_{two}$ to $P(\ket{\text{exactly two 1s}}^n_C)$.}
    \label{ngraph}
\end{figure}

As seen in Figure~\ref{ngraph}, the probability of measuring $\ket{0}^n_C$ and the entanglement signatures each converge to $\frac{1}{2}$ as $n$ increases, for both maximally entangled cases. The entanglement signature probabilities are always greater for GHZ states than for W states, and as $n$ increases so does the signature probability. This suggests that the entanglement signature probabilities are related to the amount of entanglement in the test states.

%%%%%%%%%%%%%%%%%%%%%%%%%%%%%%%%%%%%%%%%%%%%%%
\subsection{Quantifying the amount of entanglement}\label{sec:degree}

% 2-qubit
If $\ket{A} = \ket{B}$, the two-qubit probability expressions in terms of the concurrence $C_2$ from equation \eqref{c} are
\begin{align}
    P(\ket{00}_C) &= 1 - \frac{C_2^2}{4}, \nonumber\\
    P(\ket{01}_C) &= 0, \nonumber\\
    P(\ket{10}_C) &= 0 , \nonumber\\
    P(\ket{11}_C) &= \frac{C_2^2}{4} .
\end{align}
Therefore, if the control state's probability distribution is obtained (from repeats of the c-SWAP test), these results can be used to calculate the concurrence.

\begin{figure}[t]
    \centering
    \includegraphics[width=.7\textwidth]{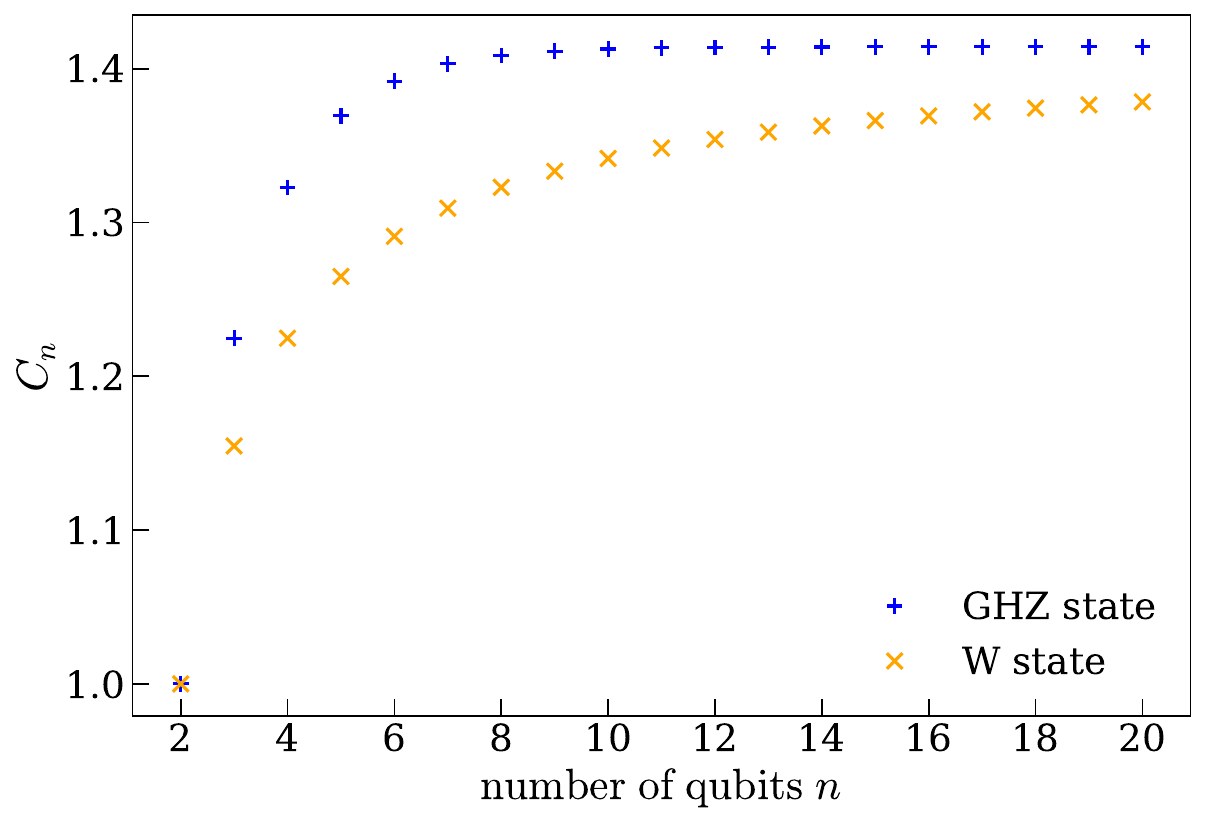}
    \caption{The value of $C_n$ from equation \eqref{concn} against number of qubits for the GHZ state from equation \eqref{defGHZ} and W state from equation \eqref{defW}.}
    \label{Wconc2}
\end{figure}

% n-qubit
Exploration of the results of the c-SWAP test for a range of example states leads us to propose a more general expression for the amount of multipartite entanglement $C_n$:
\begin{align} \label{concn}
     C_n = 2 P(\ket{\text{even no. of 1s}}_C) ^{\frac{1}{2}}
\end{align}
which is consistent with two-qubit concurrence from equation \eqref{c}.
$C_n$ therefore has a range of $0 \leq C_n \leq 2(\frac{1}{2} - \frac{1}{2^n})^\frac{1}{2}$, with the upper limit tending to $\sqrt{2}$ as $n\rightarrow \infty$. Figure.~\ref{Wconc2} shows the behaviour of this expression $C_n$ for the $\text{GHZ}_n$ and $\text{W}_n$ states. As expected, the W state has a value of $C_n$ consistently lower than that of a GHZ state. As $n$ approaches infinity the $C_n$ of both W states and GHZ states tends to $\sqrt{2}$, but at a lower rate (as a function of $n$) in the W case.

As previously discussed, a valid measure of  entanglement must satisfy the equation \eqref{entanglementdegree} for any state,
i.e.:
\begin{align}
    C_n (\ket{\psi}) \geq \sum_{j=1} p_j C_{n-1}(\ket{\psi}_j) 
\nonumber \end{align}
for any $\ket{\psi}$.
It is trivial to prove that this condition is satisfied for all W-like states (the probability expressions of which are shown in Appendix \ref{genGHZ}), and of course all GHZ-like states (because measuring a single qubit destroys all GHZ-like entanglement). From the results in \ref{3full} we have shown computationally that this condition is satisfied for any 3-qubit pure state, and we thus conjecture that it is true for any $n$-qubit pure state.

%%%%%%%%%%%%%%%%%%%%%%%%%%%%%%%%%%%%%%%%%%%%%%
\subsection{Efficiency}\label{sec:efficiency}

\begin{figure}[t]
\centering
        \includegraphics[width=.7\textwidth]{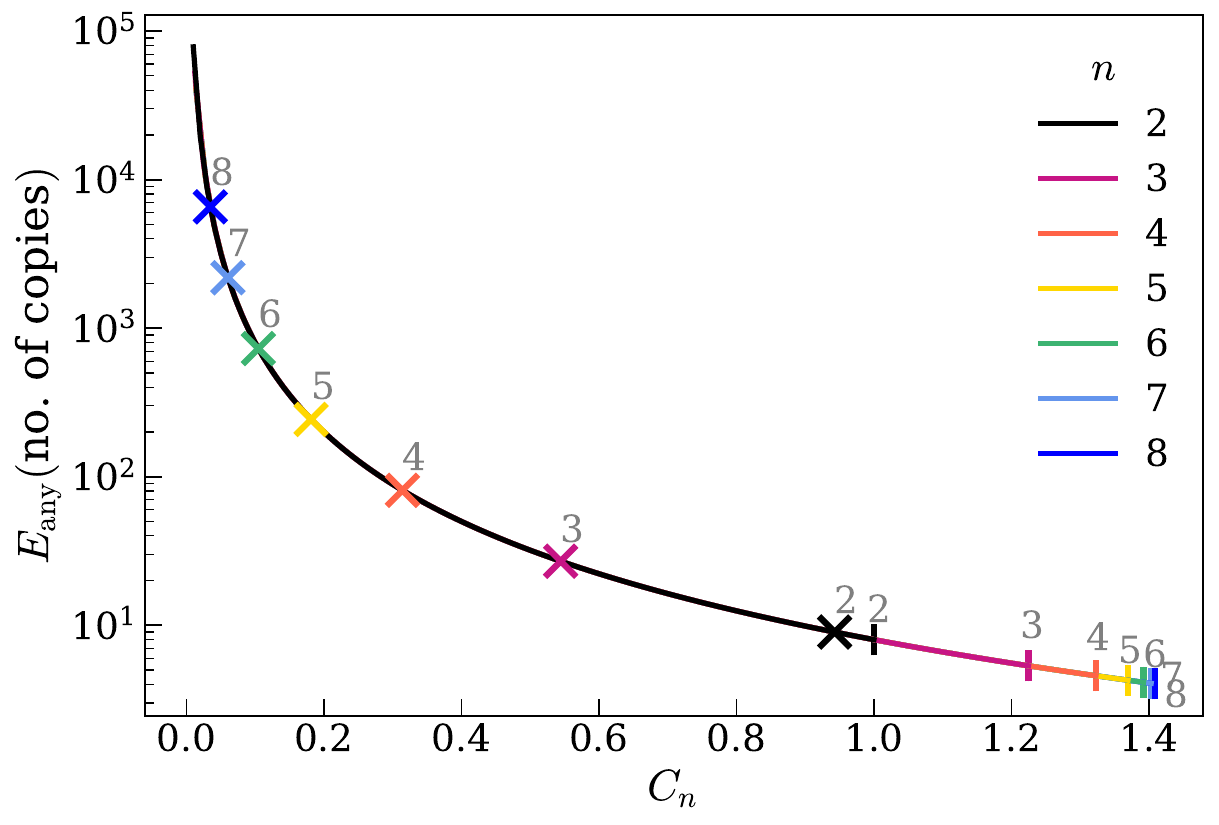}
        \caption{The expected number of copies against the amount of entanglement $C_n$ to find any entanglement in $n$-qubit test states. The vertical line segments show the upper limit of $C_n$ for each $n$. The crosses show the crossover points where the number of copies equal $3^n$, the quantum state tomography minimum scaling. Therefore those values of $C_n$ between each coloured cross and the vertical line segment of the same colour represent the regime for each $n$ (colour) under which the c-SWAP test is more favourable than quantum state tomography.}
        \label{effanyC}
\end{figure}

The resource we will be considering as a indicator of efficiency is the number of copies of the test state required for the test. If the operator of the test is only interested in detecting some entanglement in the test state (without obtaining knowledge of the amount, or whether it is genuine $n$-qubit entanglement), only one entanglement signature needs to be detected. It is straightforward to calculate the average number of measurements on the control state, and therefore the average number of copies, required to reveal the first entanglement signature. The expected number of copies is simply
\begin{align}
     E_{\text{any}}(\text{no. of copies}) = \frac{2}{P(\ket{\text{even no. of 1s}}_C)} .
\nonumber \end{align}
Therefore, the less entangled the state, the lower the entanglement signature probability, and the higher the number of measurements required. This can be illustrated in terms of proposed measure of entanglement $C_n$ from equation \eqref{concn}:
\begin{align}
     E_{\text{any}}(\text{no. of copies}) = \frac{8}{C_n^2}
\nonumber \end{align}
shown in Figure \ref{effanyC}. This expression depends on $n$ only through $C_n$, which for a maximally entangled state increases with increasing $n$ from a value of one for $n=2$ towards an upper bound of $\sqrt{2}$. Thus, for maximally entangled states, entanglement can be detected on average with eight copies or fewer. With increasing level of entanglement, the expected number of copies decreases at a rate inversely proportional to the square of $C_n$; as such, there is a large range of $C_n$ for which the expected number of copies is reasonably low.

Also plotted in Figure \ref{effanyC} are the values of $E(Y) = 3^n$, the minimum number of copies required for quantum state tomography. This figure therefore illustrates the range of $C_n$ of a given $n$-qubit state for which the c-SWAP test requires less copies (on average) than quantum state tomography: $(\frac{8}{3^n})^{\frac{1}{2}} < C_n \leq 2 (\frac{1}{2} - \frac{1}{2^n})^{\frac{1}{2}}$ (the upper bound of which is $C_n$'s absolute maximum). For example, $0.31 < C_4 \leq 1.32$ for 4-qubit states and $0.18 < C_5 \leq 1.37$ for 5-qubit states. When evidencing entanglement, there is a large regime in which the c-SWAP test outperforms quantum state tomography in terms of required number of copies. This is especially true for large systems, where for almost any amount of entanglement the c-SWAP test would be more suited than quantum state tomography.

\begin{figure}[t]
    \begin{subfigure}[b]{0.5\textwidth}
        \centering
        \includegraphics[width=1.\linewidth]{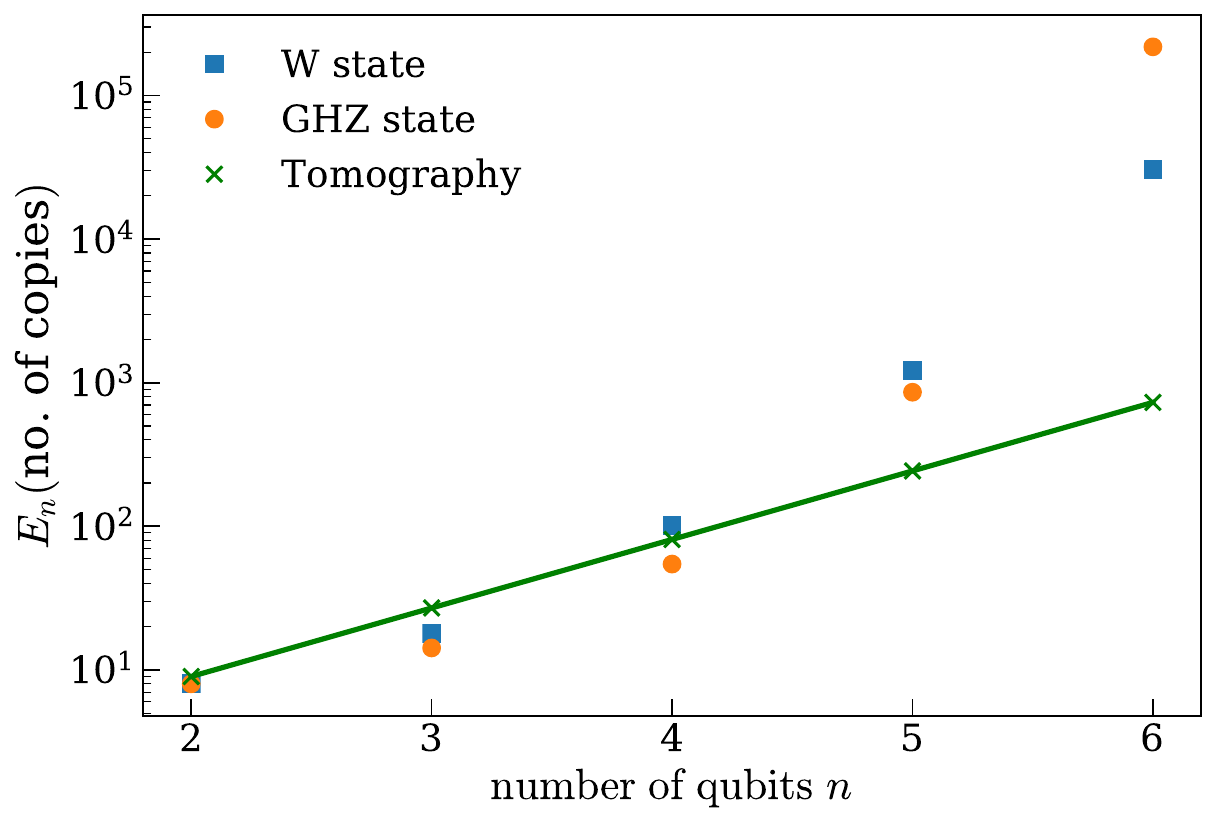}
        \caption{}
        \label{maxeff}
    \end{subfigure}\hfill
    \begin{subfigure}[b]{0.5\textwidth}
        \centering
        \includegraphics[width=1.\textwidth]{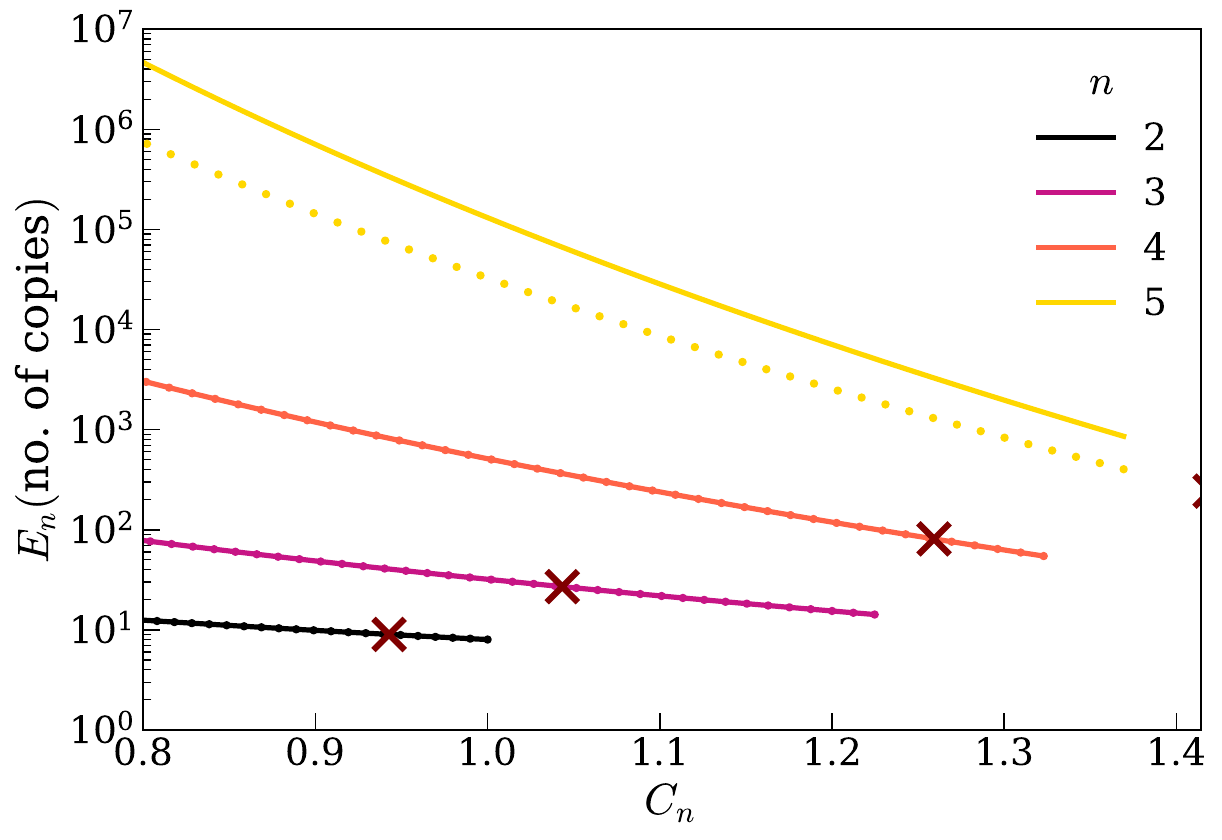}
        \caption{}
        \label{nlog}
    \end{subfigure}\hfill
    \caption{Expected number of copies needed to find genuine $n$-qubit entanglement. (a) shows test states $\ket{A} = \ket{\text{GHZ}_n} = \frac{1}{\sqrt{2}} ( \ket{0}^n + \ket{1}^n)$ and $\ket{A} = \ket{\text{W}_n} = \frac{1}{\sqrt{n}} \sum^n_{i=1} \ket{0...1_i...0}$. Also the minimum scaling for quantum state tomography, $E(\text{no. of copies}) = 3^n$. (b) shows general GHZ-like (continuous line) and W-like states (dotted line) in terms of the amount of entanglement $C_n$. The greater $n$, the greater $E_n(\text{no. of copies})$. Also shown are the crossover points (marked with crosses) for minimum scaling for quantum state tomography, $E(\text{no. of copies}) = 3^n$, such that values of $C_n$ right of the tomography plots give numbers of copies less than $3^n$.}
\end{figure}

However, if knowledge that the test state is genuinely $n$-qubit entangled is required then the required number of copies increases. Instead of detecting any one entanglement signature, one more than the total number of entanglement signatures for the $(n-1)$-qubit case must be observed. Therefore the expected number of copies are $E_n(\text{no. of copies}) = 2 [\frac{1}{2} E_{\text{any}}(\text{no. of copies})]^{x(n)}$. Example values for $x(n)$ are:
\begin{align}
     x(n)[\text{GHZ-like}] &= 2^{n-2} , \nonumber \\
     x(n)[\text{W-like}] &= \frac{1}{2}(n-1)(n-2)+1
\nonumber \end{align}
and the respective plots of $E_n(\text{no. of copies})$ are shown in Figure \ref{nlog}. The GHZ state and W state cases are shown in Figure \ref{maxeff}. The scaling with both $n$ and $C_n$ is not favourable. However, the values of $E(\text{no. of copies})$ for quantum state tomography have again been plotted and there is a regime where the number of copies required for the c-SWAP test are less than $3^n$, for states with high entanglement and less than five qubits. Therefore, carrying out this more detailed c-SWAP test is still favourable for highly entangled small systems.

%------------------------------------------------------------------------%
\section{Robustness against errors}\label{sec:errors}

To examine the robustness of the c-SWAP test we consider a range of possible errors. Clearly, it is possible that the pure state supplied is not exactly as expected. One typical example of this is that errors in the test and copy state could occur as errors in their existing non-zero amplitudes, which will be referred to as \emph{unbalanced}. Another typical error could be an additional non-zero amplitude introduced into the state, referred to as \emph{corrupted}.
Furthermore, a quantum state can also interact (entangle) with its environment and through this suffers a level of decoherence. For example, dephasing, energy dissipation, and scattering all cause decoherence, which from an ensemble perspective introduces mixture (and non-zero entropy). From a state perspective, errors in the state amplitudes arise \cite{nielsen}.
In this example it may be that only the copy state contains error and so the test state and copy state are not equivalent, which we refer to as \emph{unequal}, as would be expected from sampling a mixed ensemble. Here we also investigate an example unequal case, where the copy state is unbalanced but the test state is not.

%%%%%%%%%%%%%%%%%%%%%%%%%%%%%%%%%%%%%%%%%%%%%%
\subsection{Unbalanced}
\begin{figure}[t]
    \begin{subfigure}[b]{0.5\textwidth}
        \centering
        \includegraphics[width=1.\textwidth]{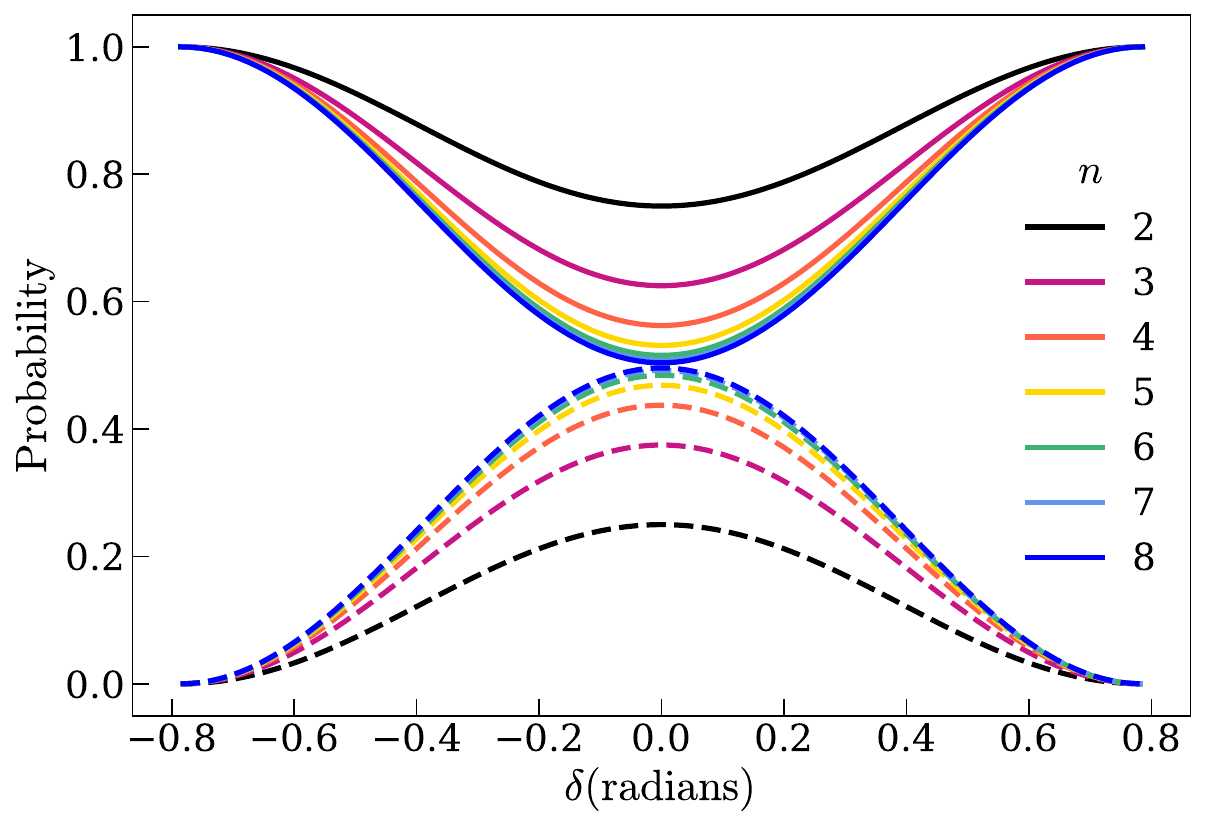}
        \caption{\label{unbalancedGHZn}}
    \end{subfigure}
    \begin{subfigure}[b]{0.5\textwidth}
        \centering
        \includegraphics[width=1.\textwidth]{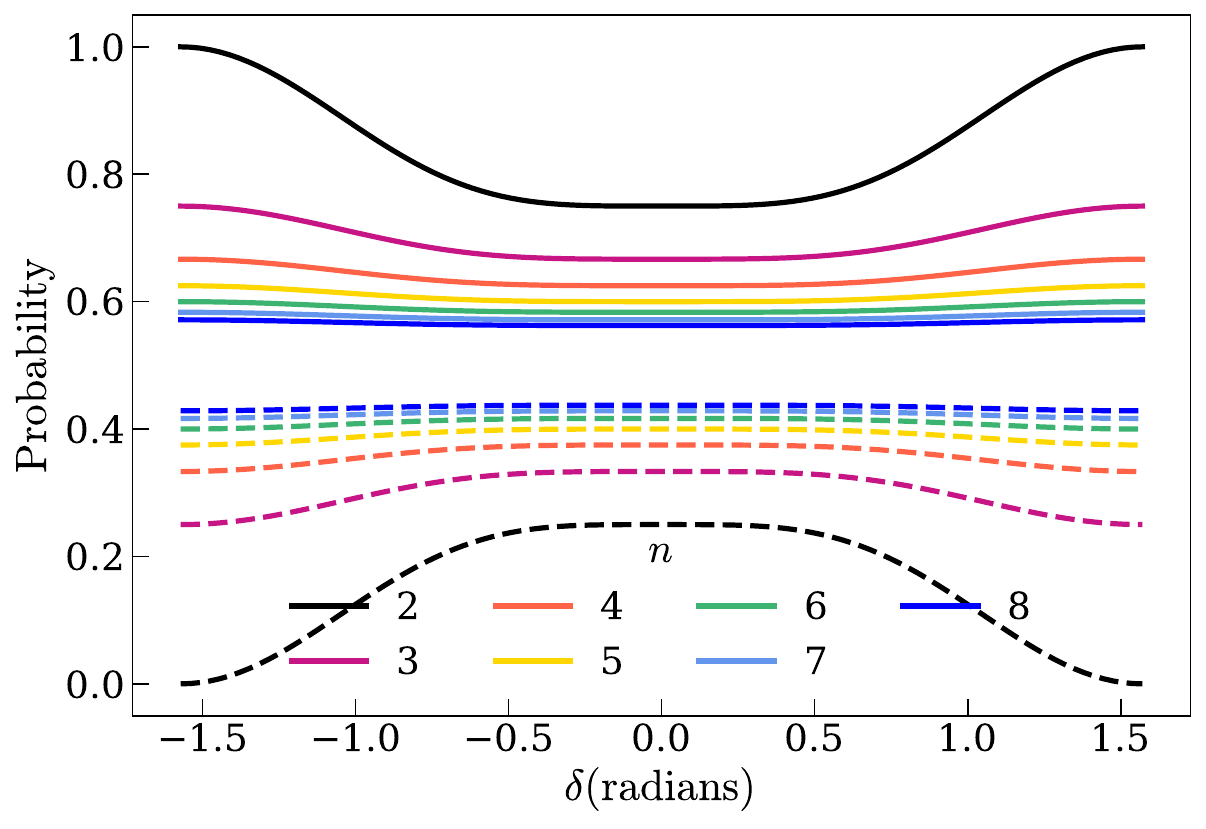}
        \caption{}
        \label{unbalancedWn}
    \end{subfigure}
    \caption{Unbalanced states with various numbers of qubits $n$, where the continuous line is $P(\ket{0}^n)$. (a) shows the probability results against $\delta$ for unbalanced GHZ test states from equation \eqref{un1}. The dashed line denotes $P(\ket{\text{even no. of 1s}}_C)$. (b) shows the results for unbalanced W test states from equation \eqref{un2}. The dashed line denotes $P(\ket{\text{exactly two 1s}}_C)$.}
\end{figure}

Our first example of error is to vary the amplitudes of otherwise maximally entangled states. Consider an
unbalanced $n$-qubit GHZ state:
\begin{align} \label{un1}
    \ket{A} = \ket{B} = \sin\left(\frac{\pi}{4} + \delta\right) \ket{0}^n + \cos\left(\frac{\pi}{4} + \delta\right) \ket{1}^n .
\end{align}
The c-SWAP test results therefore are
\begin{align}
    P(\ket{0}^n_C) &= \left[ \frac{1}{2} + \frac{1}{2^n} \right] + \left(2 - \frac{4}{2^n}\right)\cos^2\delta\sin^2\delta , \nonumber\\
    P(\ket{\text{even no. of 1s}}_C) &= \left[ \frac{1}{2} - \frac{1}{2^n} \right] - \left(2 - \frac{4}{2^n}\right)\cos^2\delta\sin^2\delta
\nonumber \end{align}
which are shown in Figure \ref{unbalancedGHZn} (as well as the unbalanced Bell state which is the $n=2$ case). The error introduced by a small non-zero value of $\delta$ is $\Delta = (2 - \frac{4}{2^n}) \delta^2$. The $n$ dependence goes to zero exponentially, so the leading order is independent of $n$. Therefore the error is approximately $2\delta^2$ for small delta and so in this case the test is robust, by which we mean that the leading order is $\delta^2$, as opposed to $\delta$.

An unbalanced $\text{GHZ}_3$ state can replicate the probability results given by a $\text{W}_3$ state. This happens in the above parametisation when $\delta = \pm \left(\sin^{-1} \sqrt{\frac{2}{3}} - \frac{\pi}{4}\right) \approx \pm 0.17$, i.e.
\begin{align}
    \ket{A} &= \ket{B} = \sqrt{\frac{2}{3}} \ket{000} + \sqrt{\frac{1}{3}} \ket{111} \nonumber \\
    \text{or} \ket{A} &= \ket{B} = \sqrt{\frac{1}{3}} \ket{000} + \sqrt{\frac{2}{3}} \ket{111} .
\nonumber \end{align}
This requires amplitude percentage errors of $15\%$ and $18\%$, a large margin of error. If necessary this uncertainty can be overcome by measuring one qubit and then applying the two-qubit c-SWAP test to the remaining state, to detect any remaining entanglement. The result would always be zero for an unbalanced $\text{GHZ}_3$ but not for a $\text{W}_3$ state. This `mimic' case is only possible with three-qubit states.

Consider an unbalanced W state with error introduced to one amplitude and the compensating error spread across the remaining amplitudes:
\begin{align} \label{un2}
    \ket{A} = \ket{B} = \sqrt{\frac{1}{n}}\cos\delta \ket{00...01} + \sqrt{\frac{1}{n-1} - \frac{1}{n(n-1)} \cos^2\delta} \: \sum^n_{j=2} \ket{0...1_j...0} .
\end{align}
This gives
\begin{align}
    P(\ket{0}^n_C) &= \left[ \frac{1}{2} + \frac{1}{2n} \right] - \frac{1}{2n^2(n-1)} \sin^2\delta \: (4(n-1) + (n-2)\sin^2\delta) \nonumber \\
    P(\ket{\text{exactly two 1s}}_C) &= \left[ \frac{1}{2} - \frac{1}{2n} \right] + \frac{1}{2n^2(n-1)} \sin^2\delta \: (4(n-1) + (n-2)\sin^2\delta) \nonumber
\end{align}
shown in Figure \ref{unbalancedWn}. For small $\delta$, the error $\Delta = \frac{2}{n^2}\delta^2$. The $n$-dependence tends to zero with increasing $n$. Unlike the GHZ case, there is no term independent of $n$ and so for large $n$ there is very little variation in probability for any $\delta$.

%%%%%%%%%%%%%%%%%%%%%%%%%%%%%%%%%%%%%%%%%%%%%%
\subsection{Unequal} \label{sec:unequal}

While the c-SWAP test requires two copies of the test state, it may be that the two generated states are not equivalent ($\ket{A} \neq \ket{B}$) if these are drawn from a mixed ensemble.

\begin{figure}[t]    
    \begin{subfigure}[b]{0.5\textwidth}
        \centering
        \includegraphics[width=1.\textwidth]{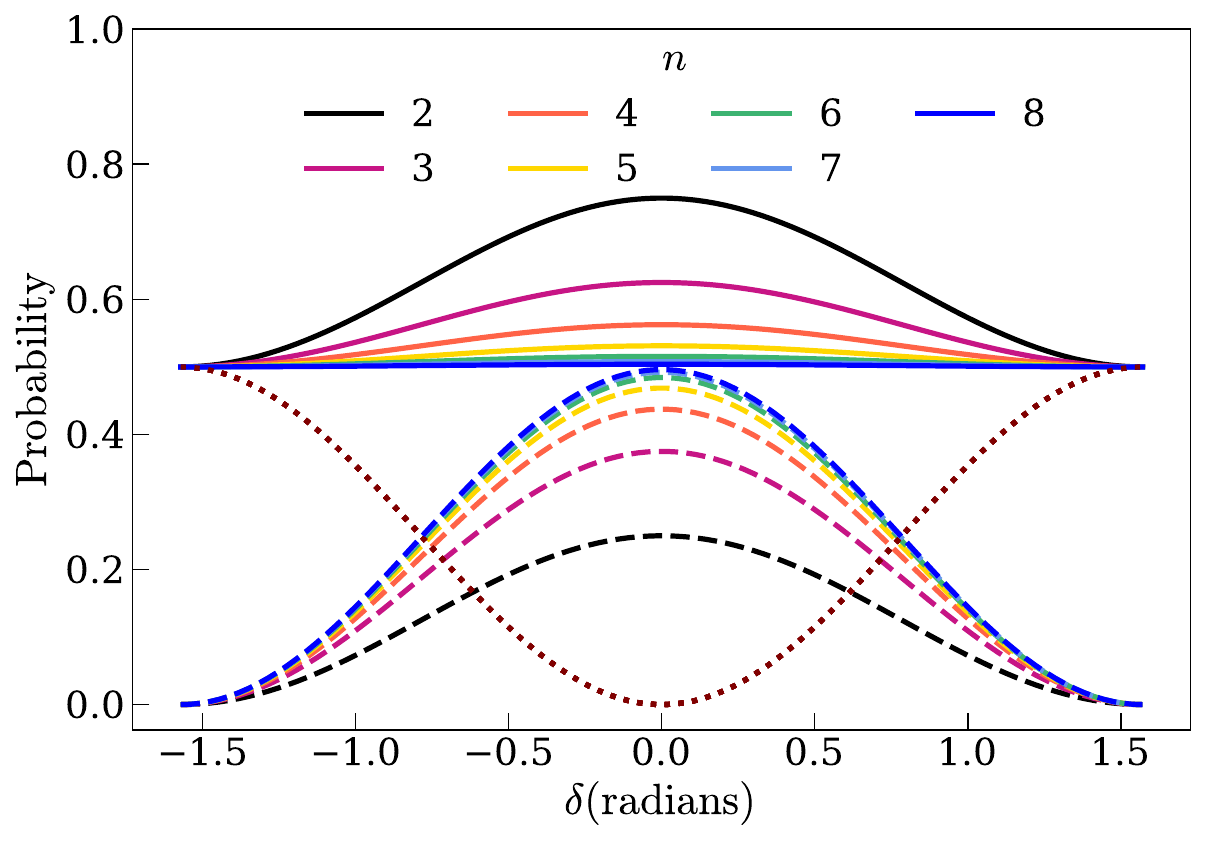}
        \caption{}
        \label{unequalGHZn}
    \end{subfigure}\hfill
    \begin{subfigure}[b]{0.5\textwidth}
        \centering
        \includegraphics[width=1.\textwidth]{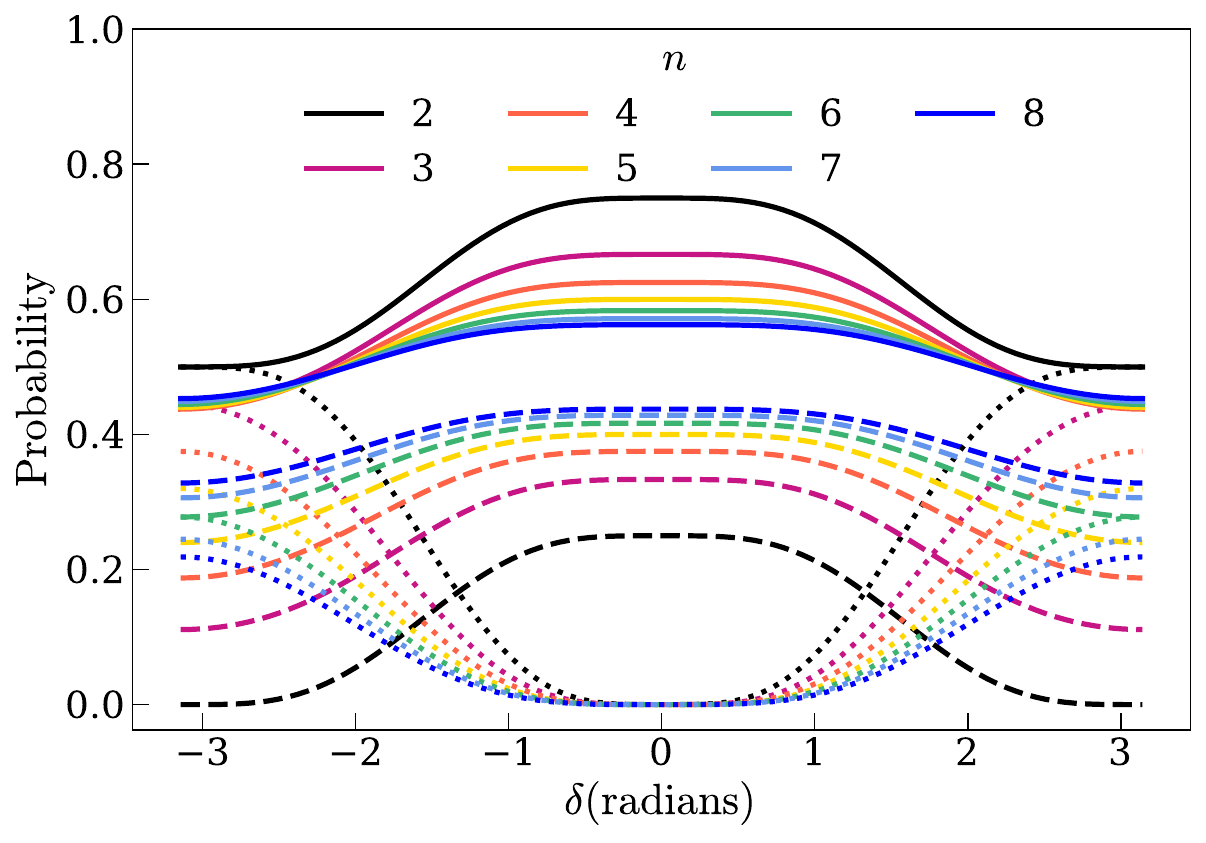}
        \caption{}
        \label{unequalWn}
    \end{subfigure}\hfill
    \caption{Unequal states for various numbers of qubits $n$. The continuous line denotes $P(\ket{0}^n)$. (a) shows the probability results against $\delta$ for inequivalent copy states where one is a GHZ state and the other is equation \eqref{une1}. The dashed line denotes $P(\ket{\text{even no. of 1s}}_C)$ and the dotted line denotes $P(\ket{\text{odd no. of 1s}}_C)$ (which is the same for all $n$). (b) shows the results for one W state and copy state equation \eqref{une2}. The dashed line denotes $P(\ket{\text{exactly two 1s}}_C)$ and the dotted line denotes $P(\ket{\text{exactly one 1}})$.}
\end{figure}

% unequal GHZ n
Consider the error case $\ket{A} = \ket{\text{GHZ}_n}$ (or $\ket{A} = \ket{\Phi^+}$ in the $n=2$ case) and
\begin{align} \label{une1}
    \ket{B} = \sin\left(\frac{\pi}{4} + \delta\right) \ket{0}^n + \cos\left(\frac{\pi}{4} + \delta\right) \ket{1}^n ,
\end{align}
giving:
\begin{align}
     P(\ket{0}^n_C) &= \left[ \frac{1}{2} + \frac{1}{2^n} \right] - \frac{1}{2^n}\sin^2\delta , \nonumber\\
     P(\ket{\text{odd no. of 1s}}_C) &= \frac{1}{2} \sin^2\delta , \nonumber\\
     P(\ket{\text{even no. of 1s}}_C) &= \left[\frac{1}{2} - \frac{1}{2^n} \right] - \left(\frac{1}{2} - \frac{1}{2^n}\right) \sin^2\delta
\nonumber \end{align}
shown in Figure \ref{unequalGHZn}. Any measurement of an odd number of $\ket{1}$s in the control therefore demonstrates that the test state and copy state are not equivalent.

This is similar to the c-SWAP test for equivalence from Section \ref{sec:equalSWAP}, where any measurement of $\ket{1}$ in the control qubit demonstrates the two test states are inequivalent. The equivalency c-SWAP test applied to the same test states from equation \eqref{une1} that we have just considered for the entanglement c-SWAP gives the resulting probability of measuring $\ket{1}_C$ (and therefore certainty of inequivalence) as:
\begin{align}
    P(\ket{1}_C) = \frac{1}{2}\cos\delta\sin\delta \nonumber
\end{align}
which with small $\delta$ approximates to $\Delta_1 = \frac{1}{2}\delta$. This is in fact less robust than the entanglement test, the errors for which are all second order: for small $\delta$, the errors satisfy $\Delta_0 = \frac{1}{2^n} \delta^2$, $\Delta_{\text{odd}} = \frac{1}{2} \delta^2$, and $\Delta_{\text{even}} = (\frac{1}{2} - \frac{1}{2^n})\delta^2$.

Interestingly, the unequal states signature probability, $P(\ket{\text{odd no. of 1s}}_C)$, has no $n$-dependence. Where it is present, the $n$-dependence again is confined to the coefficients and the error tend to zero exponentially with $n$. As its error has no term independent of $n$, $P(\ket{0}^n_C)$ tends to $\frac{1}{2}$ for large systems and so cannot be used as as indicator of large error. $P(\ket{\text{odd no. of 1s}}_C)$ and $P(\ket{\text{even no. of 1s}}_C)$ however always vary with $\delta$.

% unequal W_n
Similarly, $\ket{A} \neq \ket{B}$ for the W case can be investigated where $\ket{A} = \ket{\text{W}_n}$ and
\begin{align} \label{une2}
    \ket{B} = \sqrt{\frac{1}{n}}\cos\delta \ket{00...01} + \sqrt{\frac{1}{n-1} - \frac{1}{n(n-1)} \cos^2\delta} \: \sum^n_{j=2} \ket{0...1_j...0} .
\end{align}
This gives:
\begin{align}
    P(\ket{0}^n_C) &= \left[ \frac{1}{2} + \frac{1}{2n} \right] - \frac{n-1}{4n^2} \left( \cos^2\delta + 1 - 2 \cos\delta \: \sqrt{1 + \frac{\sin^2\delta}{n-1}} \right) \nonumber \\
    P(\ket{\text{exactly one 1s}}_C) &= \frac{n-1}{2n^2} \left( \cos^2\delta + 1 - 2 \cos\delta \: \sqrt{1 + \frac{\sin^2\delta}{n-1}} \right) \nonumber \\
    P(\ket{\text{exactly two 1s}}_C) &= \left[ \frac{1}{2} - \frac{1}{2n} \right] - \frac{n-1}{4n^2} \left( \cos^2\delta + 1 - 2 \cos\delta \: \sqrt{1 + \frac{\sin^2\delta}{n-1}} \right) \nonumber
\end{align}
shown in Figure \ref{unequalWn}. For small $\delta$, $\Delta_0 = \Delta_{\text{two 1s}} = \frac{1}{4n^2} \delta^2$ and $\Delta_{\text{one 1}} = \frac{1}{2n^2} \delta^2$. The leading order errors vanish inversely with increasing $n^2$.

%%%%%%%%%%%%%%%%%%%%%%%%%%%%%%%%%%%%%%%%%%%%%%
\subsection{Corrupted}

\begin{figure}[t]
    \begin{subfigure}[b]{0.5\textwidth}
        \centering
        \includegraphics[width=1.\textwidth]{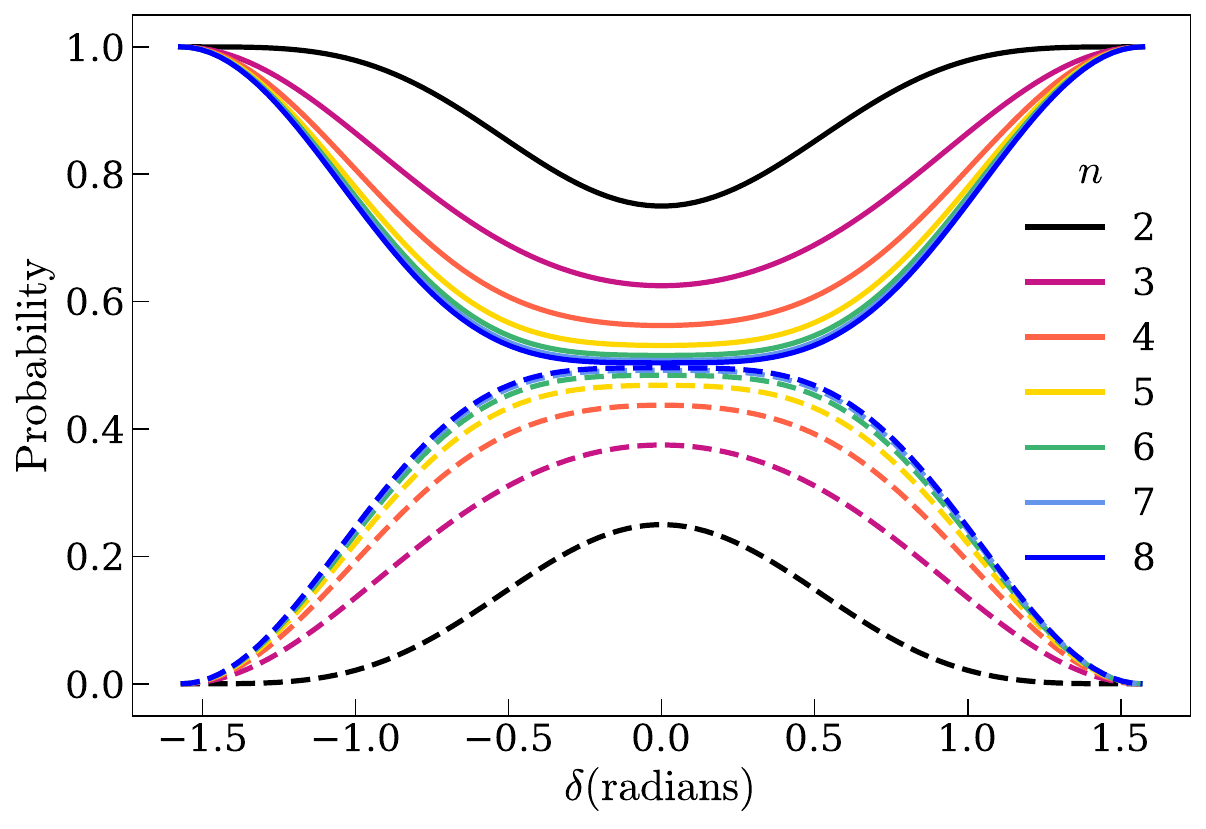}
        \caption{}
        \label{GHZc}
    \end{subfigure}\hfill
    \begin{subfigure}[b]{0.5\textwidth}
        \centering
        \includegraphics[width=1.\textwidth]{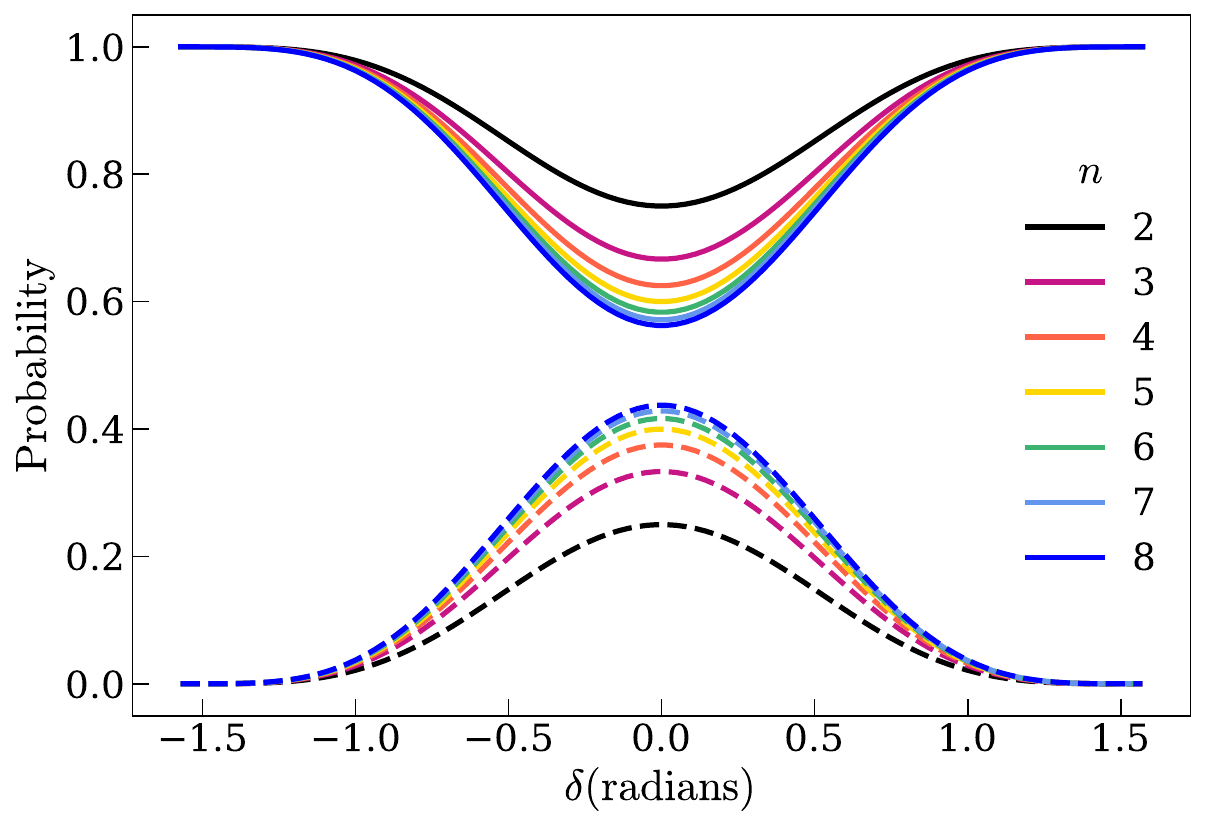}
        \caption{}
        \label{Wc}
    \end{subfigure}\hfill
    \caption{Corrupted states with various numbers of qubits $n$. The continuous line denotes $P(\ket{0}^n)$. (a) shows the probability results against $\delta$ for states equation \eqref{cor1}. The dotted line denotes $P(\ket{\text{even no. of 1s}}_C)$. (b) shows the results for states equation \eqref{cor2}. The dotted line denotes $P(\ket{\text{exactly two 1s}}_C)$.}
\end{figure}

Another source of error is to `corrupt' an entangled state by introducing an additional non-zero amplitude. The following cases assume the test state and copy state are exact copies.
% GHZ n
The case
\begin{align} \label{cor1}
    \ket{A} = \ket{B} = \cos\delta\ket{\text{GHZ}_n} + \sin\delta\ket{0...1}
\end{align}
gives:
\begin{align}
    P(\ket{0}^n_C) &= \left[ \frac{1}{2} + \frac{1}{2^n} \right] + \frac{1}{2^n}{\sin^2\delta}  \: (2 + (2^{n-1} - 3)\sin^2\delta) , \nonumber\\
    P(\ket{\text{even no. of 1s}}_C) &= \left[ \frac{1}{2} - \frac{1}{2^n} \right] - \frac{1}{2^n}\sin^2\delta  \: (2 + (2^{n-1} - 3)\sin^2\delta)
\nonumber \end{align}
shown in Figure \ref{GHZc}. For small $\delta$, the error is $\Delta = \frac{2}{2^n} \delta^2$ and so again the errors tend to zero exponentially with $n$.
Unlike the other GHZ examples, the individual signature probabilities are not equal to one another. The probabilities for states ending with $\ket{1}$s and those ending with $\ket{0}$s have different values, with the former independent of $n$. This is due to the final $\ket{1}$ in the additional state and alternative `extra' states give different individual probabilities.

% W n
Unlike the GHZ case, the corrupted W state results depend on which state is added. For example if
\begin{align} \label{cor2}
    \ket{A} = \ket{B} = \cos\delta\ket{\text{W}_n} + \sin\delta\ket{0}^n
\end{align}
then
\begin{align}
    P(\ket{0}^n_C) &= \left[ \frac{1}{2} + \frac{1}{2n} \right] + \frac{n - 1}{2n} \sin^2\delta \: (2 - \sin^2\delta) , \nonumber \\
    P(\ket{\text{exactly two 1s}}_C) &= \left[ \frac{1}{2} - \frac{1}{2n} \right] - \frac{n-1}{2n} \sin^2\delta \: (2 - \sin^2\delta)
\nonumber \end{align}
shown in Figure \ref{Wc}. For small $\delta$ the errors are $\Delta = (1 - \frac{1}{n}) \delta^2$. The leading order is therefore independent of $n$, with the $n$ dependence tending to zero inversely with $n$.

%------------------------------------------------------------------------%
\section{Summary and conclusions}\label{sec:conc}

Entanglement is essential for quantum information processes such as quantum teleportation. The controlled SWAP test is a proposed method to detect and quantify entanglement for any $n$-qubit pure state. We have investigated when it is practical to use and potentially more efficient than quantum state tomography.  
In terms of the required number of copies of the test state, the c-SWAP test for entanglement is more efficient for detecting entanglement for almost all states except two-qubit states.
The average number of copies required to detect the presence of entanglement can be as low as four for larger high fidelity maximally entangled states.  For perfect Bell states, it typically requires eight copies, and for states that are not maximally entangled it can rise to a thousand or more, with more copies required the less entangled the state is. 
The number of qubits $n$ in the state has considerably less effect on this value, which in fact decreases with increasing $n$, and so the number of copies scales extremely well with system size.
However, the short gate sequence required for the c-SWAP test for entanglement is more complicated than the single qubit rotations typically \cite{tomog2} required for quantum state tomography.
Especially for linear optics \cite{KLM}, additional ancillary resources are needed if a probabilistic set up is used, to teleport in the gates, for example \cite{Gottesman2,Bennett}.  In other settings, for example Rydberg atomic qubits \cite{rydberg}, the long range interactions may facilitate the multi-qubit Toffoli gates \cite{toffoli}, making the c-SWAP test more viable.

The test is also able to distinguish classes of entanglement in almost all cases (there is one case of three qubits for which the test is fooled, but this requires 18\% amplitude error in the generated state). The results from a two-qubit test state are directly related to the concurrence of the state. Further, a multipartite measure of entanglement has been constructed that is given by any state's c-SWAP test results.
Hence, the c-SWAP test can be used to estimate amount of entanglement in the test state.
Detecting genuine $n$-qubit entanglement is more involved, and scales far less favourably with both $n$ and the amount of entanglement, but is feasible for small numbers of qubits.
Even though the test is capable of quantifying the amount of entanglement in the state, it is only suited to achieving this with small highly entangled systems.

The suitability of the entanglement SWAP test for experimental implementation is highlighted by the fact that various typical small deviations from ideal states all give second order errors for any number of qubits.
This favourable error dependence allows the entanglement to be estimated accurately in a practical set up.

Future work, beyond the results reported here, could investigate in more detail the application of the controlled SWAP test to mixed states and other types of entangled states (qudits, coherent states). This would expand on our consideration of corrupted, unbalanced,
and unequal test and copy states, and provide further information to support the practical application of the c-SWAP test for entanglement.

\paragraph{Acknowledgements}
We thank Adam Callison for help creating the figures. SF is supported by a UK EPSRC funded DTG studentship.

\bibliography{cswap.bib}

\appendix
\addtocontents{toc}{\fixappendix}
\appendixpage

\section{Two-qubit probability results}\label{2full}
For completely general test states from equation \eqref{super2} where $\ket{A} \neq \ket{B}$:
\begin{align}
    P(\ket{00}_C) &= \frac{1}{4} [ 4 ( {A_{00}}^2 {B_{00}}^2 + {A_{01}}^2 {B_{01}}^2 + {A_{10}}^2 {B_{10}}^2 + {A_{11}}^2 {B_{11}}^2 ) \nonumber\\
    & \; \; \; \; + 2 ( A_{00} B_{01} + A_{01} B_{00} )^2 + 2 ( A_{00} B_{10} + A_{10} B_{00} )^2 \nonumber\\
    & \; \; \; \; + 2 ( A_{01} B_{11} + A_{11} B_{01} )^2 + 2 ( A_{10} B_{11} + A_{11} B_{10} )^2 \nonumber\\
    & \; \; \; \; + ( A_{00} B_{11} + A_{01} B_{10} + A_{10} B_{01} + A_{11} B_{00} )^2 ] , \nonumber\\
    P(\ket{01}_C) &= \frac{1}{4} [ 2 ( A_{00} B_{01} - A_{01} B_{00} )^2 + 2 ( A_{10} B_{11} - A_{11} B_{10} )^2 \nonumber\\
    & \; \; \; \; + ( A_{00} B_{11} - A_{01} B_{10} + A_{10} B_{01} - A_{11} B_{00} )^2 ] , \nonumber\\
    P(\ket{10}_C) &= \frac{1}{4} [ 2 ( A_{00} B_{10} - A_{10} B_{00} )^2 + 2 ( A_{01} B_{11} - A_{11} B_{01} )^2 \nonumber\\
    & \; \; \; \; + ( A_{00} B_{11} + A_{01} B_{10} - A_{10} B_{01} - A_{11} B_{00} )^2 ] , \nonumber\\
    P(\ket{11}_C) &= \frac{1}{4} ( A_{00} B_{11} - A_{01} B_{10} - A_{10} B_{01} + A_{11} B_{00} )^2
\end{align}
If $\ket{A} = \ket{B}$:
\begin{align}
    P(\ket{00}_C) &= 1 - ( A_{00} A_{11} - A_{01} A_{10} )^2 \nonumber\\
    &= 1 - \frac{1}{4} C_2^2 , \nonumber\\
    P(\ket{01}_C) &= 0 , \nonumber\\
    P(\ket{10}_C) &= 0 , \nonumber\\
     P(\ket{11}_C) &= ( A_{00} A_{11} - A_{01} A_{10} )^2 \nonumber\\
    &= \frac{1}{4} C_2^2
\end{align}
where $C_2$ is the concurrence.

\section{Three-qubit probability results}\label{3full}
For completely general test states where $\ket{A} = \ket{B}$:
\begin{align}
    P(\ket{000}_C) &= \frac{1}{2} [ 2( {A_{000}}^4 + {A_{001}}^4 + {A_{010}}^4 + {A_{011}}^4 \nonumber\\
    & \; \; \; \;  + {A_{100}}^4 + {A_{101}}^4 + {A_{110}}^4 + {A_{111}}^4 ) \nonumber\\
    & \; \; \; \;  + 4  {A_{000}}^2 ( {A_{001}}^2 + {A_{010}}^2 + {A_{100}}^2 ) \nonumber\\
    & \; \; \; \;  + 4  {A_{011}}^2 ( {A_{001}}^2 + {A_{010}}^2 + {A_{111}}^2 ) \nonumber\\
    & \; \; \; \;  + 4  {A_{101}}^2 ( {A_{001}}^2 + {A_{100}}^2 + {A_{111}}^2 ) \nonumber\\
    & \; \; \; \;  + 4  {A_{110}}^2 ( {A_{010}}^2 + {A_{100}}^2 + {A_{111}}^2 ) \nonumber\\
    & \; \; \; \;  + 2( A_{000} A_{011} + A_{001} A_{010} )^2 \nonumber\\
    & \; \; \; \;  + 2( A_{000} A_{101} + A_{001} A_{100} )^2 \nonumber\\
    & \; \; \; \;  + 2( A_{000} A_{110} + A_{010} A_{100} )^2 \nonumber\\
    & \; \; \; \;  + 2( A_{001} A_{111} + A_{011} A_{101} )^2 \nonumber\\
    & \; \; \; \;  + 2( A_{010} A_{111} + A_{011} A_{110} )^2 \nonumber\\
    & \; \; \; \;  + 2( A_{100} A_{111} + A_{101} A_{110} )^2 \nonumber\\
    & \; \; \; \;  + ( A_{000} A_{111} + A_{001} A_{110} + A_{010} A_{101} + A_{011} A_{100} )^2 ] , \nonumber\\
    P(\ket{001}_C) &= 0 , \nonumber\\
    P(\ket{010}_C) &= 0 , \nonumber\\
    P(\ket{011}_C) &= \frac{1}{2} [ 2( A_{000} A_{011} - A_{001} A_{010} )^2 \nonumber\\
    & \; \; \; \;  + 2( A_{100} A_{111} - A_{101} A_{110} )^2 \nonumber\\
    & \; \; \; \;  + ( A_{000} A_{111} - A_{001} A_{110} - A_{010} A_{101} + A_{011} A_{100} )^2 ] , \nonumber\\
    P(\ket{100}_C) &= 0 , \nonumber\\
    P(\ket{101}_C) &= \frac{1}{2} [ 2( A_{000} A_{101} - A_{001} A_{100} )^2 \nonumber\\
    & \; \; \; \;  + 2( A_{010} A_{111} - A_{011} A_{110} )^2 \nonumber\\
    & \; \; \; \;  + ( A_{000} A_{111} - A_{001} A_{110} + A_{010} A_{101} - A_{011} A_{100} )^2 ] , \nonumber\\
    P(\ket{110}_C) &= \frac{1}{2} [ 2( A_{000} A_{110} - A_{010} A_{100} )^2 \nonumber\\
    & \; \; \; \;  + 2( A_{001} A_{111} - A_{011} A_{101} )^2 \nonumber\\
    & \; \; \; \;  + ( A_{000} A_{111} + A_{001} A_{110} - A_{010} A_{101} - A_{011} A_{100} )^2 ] , \nonumber\\
    P(\ket{111}_C) &= 0
\end{align}
For general GHZ-like test states where $\ket{A} \neq \ket{B}$, $\ket{A} = A_{000} \ket{000} + A_{111} \ket{111}$ and $\ket{B} = B_{000} \ket{000} + \ket{111} B_{111}$:
\begin{align}
    P(\ket{000}_C) &= {A_{000}}^2 {B_{000}}^2 + {A_{111}}^2 {B_{111}}^2 \nonumber\\
    & \; \; \; \;  + \frac{1}{8} ( A_{000} B_{111} + A_{111} B_{000} )^2 , \nonumber\\
    P(\ket{001}_C) &= \frac{1}{8} ( A_{000} B_{111} - A_{111} B_{000} )^2 \nonumber\\
    &= P(\ket{010}_C) = P(\ket{100}_C) = P(\ket{111}_C) , \nonumber\\
    P(\ket{011}_C) &= \frac{1}{8} ( A_{000} B_{111} + A_{111} B_{000} )^2 \nonumber\\
    &= P(\ket{101}_C) = P(\ket{110}_C)
\end{align}
For general W-like test states where $\ket{A} \neq \ket{B}$, $\ket{A} = A_{001} \ket{001} + A_{010} \ket{010} + A_{100} \ket{100}$ and $\ket{B} = B_{001} \ket{001} + B_{010} \ket{010} + B_{100} \ket{100}$:
\begin{align}
     P(\ket{000}_C) &= {A_{001}}^2{B_{001}}^2 + {A_{010}}^2{B_{010}}^2 + {A_{100}}^2{B_{100}}^2 \nonumber\\
     & \; \; \; \;  + \frac{1}{4}[(A_{001}B_{010} + A_{010}B_{001})^2 + (A_{001}B_{100} + A_{100}B_{001})^2 \nonumber\\
     & \; \; \; \;  + (A_{010}B_{100} + A_{100}B_{010})^2] , \nonumber\\
     P(\ket{001}_C) &= \frac{1}{4}[(A_{001}B_{010} - A_{010}B_{001})^2 \nonumber\\
     & \; \; \; \;  + (A_{001}B_{100} - A_{100}B_{001})^2] , \nonumber\\
     P(\ket{010}_C) &= \frac{1}{4}[(A_{001}B_{010} - A_{010}B_{001})^2 \nonumber\\
     & \; \; \; \;  + (A_{010}B_{100} - A_{100}B_{010})^2] , \nonumber\\
     P(\ket{011}_C) &= \frac{1}{4}(A_{001}B_{010} + A_{010}B_{001})^2 , \nonumber\\
     P(\ket{100}_C) &= \frac{1}{4}[(A_{001}B_{100} - A_{100}B_{001})^2 \nonumber\\
     & \; \; \; \;  + (A_{010}B_{100} - A_{100}B_{010})^2] , \nonumber\\
     P(\ket{101}_C) &= \frac{1}{4}(A_{001}B_{100} + A_{100}B_{001})^2 , \nonumber\\
     P(\ket{110}_C) &= \frac{1}{4}(A_{010}B_{100} + A_{100}B_{010})^2 , \nonumber\\
     P(\ket{111}_C) &= 0 .
\end{align}

\section{$n$-qubit probability results}\label{nfull}
For general unbalanced GHZ case $\ket{A} = \ket{B} = \alpha_0 \ket{0}^n + \alpha_1 \ket{1}^n$:
\begin{align} \label{unGHZgen}
    P(\ket{0}^n_C) &= 1 - \frac{2^{n-1} -1}{2^{n-2}} \alpha_0^2 \alpha_1^2 , \nonumber\\
    P(\ket{\text{even no. of 1s}}_C) &= \frac{2^{n-1}-1}{2^{n-2}} \alpha_0^2 \alpha_1^2 .
\end{align}
For general unbalanced W case $\ket{A} = \ket{B} = 
a_1 \ket{00...1} + a_2 \sum^n_{j=2} \ket{0...1_j...0}$:
\begin{align} \label{unWgen}
    P(\ket{0}^n_C) &= 1 - (n-1) a_2^2 \left( a_1^2 + \frac{n-1}{2} a_2^2 \right) , \nonumber\\
    P(\ket{\text{exactly two 1s}}_C) &= (n-1) a_2^2 \left( a_1^2 + \frac{n-1}{2} a_2^2 \right)  .
\end{align}
For general GHZ-like states where $\ket{A} \neq \ket{B}$, $\ket{A} = \alpha_0 \ket{0}^n + \alpha_1 \ket{1}^n$ and $\ket{B} = \beta_0 \ket{0}^n + \beta_1 \ket{1}^n$:
\begin{align} \label{genGHZ}
     P(\ket{0}^n_C) &= \alpha_0^2 \beta^2_0 + \alpha_1^2 \beta^2_1 + \frac{1}{2^n} (\alpha_0 \beta_1 + \alpha_1 \beta_0)^2 , \nonumber\\
     P(\ket{\text{odd no. of 1s}}_C) &= \frac{1}{2} - (\alpha_0 \beta_0 + \alpha_1 \beta_1)^2 , \nonumber\\
     P(\ket{\text{even no. of 1s}}_C) &= \frac{2^{n-1}-1}{2^n} (\alpha_0 \beta_1 + \alpha_1 \beta_0)^2 .
\end{align}
For general W-like states where $\ket{A} \neq \ket{B}$, $\ket{A} = \sum^n_{i=1} a_i \ket{0...1_i...0}$, $\ket{B} = \sum^n_{j=1} b_j \ket{0...1_j...0}$:
\begin{align} \label{genW}
     P(\ket{0}^n_C) &= \sum^n_{i=1} \left( a_i^2 b_i^2 + \frac{1}{8} \sum^n_{j=1, j \neq i} (a_i b_j + a_j b_i)^2 \right) , \nonumber\\
     P(\ket{\text{exactly one 1}}_C) &= \frac{1}{4} \sum^n_{i=1} \sum^n_{j=1, j \neq i} (a_i b_j - a_j b_i)^2 , \nonumber\\
     P(\ket{\text{exactly two 1s}}_C) &= \sum^n_{i=1} \sum^n_{j=1, j \neq i} \frac{1}{8} (a_i b_j + a_j b_i)^2 .
\end{align}
and the individual probabilities are:

$P(\ket{0...1_i...0}_C) = \frac{1}{4} \sum^n_{j=1, j \neq i} (a_i b_j - a_j b_i)^2$, $P(\ket{0...1_i...1_j...0}_C) =  \frac{1}{4} (a_i b_j + a_j b_i)^2$.
\end{document}